\documentclass[manuscript]{acmart}
\AtBeginDocument{%
	}

\usepackage{fancyhdr}
\usepackage[ruled, linesnumbered]{algorithm2e}
\usepackage{graphicx}
\usepackage{subfigure}
\usepackage{url}
\usepackage{multirow}
\usepackage{array}
\usepackage{booktabs}
\usepackage{enumitem}
\usepackage{rotating}

\usepackage{algorithmic}

\usepackage{amsmath,amssymb,amsthm}  

\usepackage{xcolor}
\usepackage{color}

\usepackage{listings}
\usepackage{caption}

\setcopyright{acmcopyright}
\acmJournal{TOSEM}

\begin{document}
\title{Towards an Oracle for Binary Decomposition Under Compilation Variance}

\author{Ang Jia}
\author{He Jiang}
\authornote{Corresponding author.}
\author{Zhilei Ren}
\author{Xiaochen Li}
\author{Zhipeng Yang}
\author{Yaxin Duan}
\affiliation{%
	\institution{School of Software, Dalian University of Technology}
	\city{Dalian}
	\country{China}}
\email{{jiaang,jianghe,zren,xiaochen.li}@dlut.edu.cn,22417020@mail.dlut.edu.cn, dyx2377738569@163.com}

\author{Ming Fan}
\author{Ting Liu}
\affiliation{%
	\institution{MOE Key Laboratory of Intelligent Networks and Network Security, Xi'an Jiaotong University}
	\city{Xi'an}
	\country{China}}
\email{{mingfan,tingliu}@mail.xjtu.edu.cn}

\begin{abstract}
	Third-Party Library (TPL) detection, which identifies reused libraries in binary code, is critical for software security analysis. At its core, TPL detection depends on binary decomposition---the process of partitioning a monolithic binary into cohesive modules. Existing decomposition methods, whether anchor-based or clustering-based, fundamentally rely on the assumption that reused code exhibits similar function call relationships. However, this assumption is severely undermined by Function Call Graph (FCG) variations introduced by diverse compilation settings, particularly function inlining decisions that drastically alter FCG structures.
	
	In this work, we conduct the first systematic empirical study to establish the oracle for optimal binary decomposition under compilation variance. We first develop a labeling method to create precise FCG mappings on a comprehensive dataset compiled with 17 compilers, 6 optimizations, and 4 architectures; then, we identify the minimum semantic-equivalent function regions between FCG variants to derive the ground-truth decomposition. This oracle provides the first rigorous evaluation framework that quantitatively assesses decomposition algorithms under compilation variance. Using this oracle, we evaluate existing methods and expose their critical limitations: they either suffer from under-aggregation failure or over-aggregation failure. Our findings reveal that current decomposition techniques are inadequate for robust TPL detection, highlighting the urgent need for compilation-aware approaches.
\end{abstract}




\maketitle

\section{Introduction}
\label{sec:introduction}

Most software today is not developed entirely from scratch. Instead, developers rely on a range of Third-Party Libraries (TPLs) to create their applications~\cite{kula2018developers}. According to a recent report~\cite{CodeResue}, 97\% of the software contains at least one TPL. Although using TPLs helps to finish projects quicker and reduce costs, improper reuse introduces security and legal risks~\cite{gkortzis2021software}. Of the 1,703 codebases scanned in 2022, 87\% include security and operational risks~\cite{CodeResue}, where 54\% of software has license conflicts and 84\% of software contains at least one vulnerability. To make it worse, downstream software often relies on closed-source TPLs~\cite{TPL}. Security and legal risks hidden in the binaries generated by the upstream supplier may be unintentionally transferred to the downstream users.


To resolve these code-reuse-related issues, TPL reuse detection works~\cite{li2023libam, xu2021interpretation, dong2024libvdiff, tang2022libdb, zhu2022bbdetector, yang2022modx, sun2023moddiff, karande2018bcd, guo2023searching, tang2020libdx} are proposed. These detection works first decompose the TPL into modules, and then conduct similarity detection between binary modules. As pointed out by many existing works~\cite{li2023libam, xu2021interpretation, yang2022modx}, TPLs are usually wholly or partially used in binaries, thus existing works either apply anchor-based methods~\cite{li2023libam, xu2021interpretation, dong2024libvdiff, tang2022libdb} by extending anchor functions to generate modules or apply clustering-based method~\cite{yang2022modx, sun2023moddiff, karande2018bcd, guo2023searching} by using clustering algorithms to group binary functions. Regardless of these differences, they all rely on that reused code still shares similar function call relationships.

\begin{figure*}[h]
	\centering
	\subfigure[FCG of BZ2\_compressBlock in Bzip2 (compiled by GCC using O1)]{
		\begin{minipage}[t]{0.95\linewidth}
			\centering
			\includegraphics[width=0.95\textwidth]{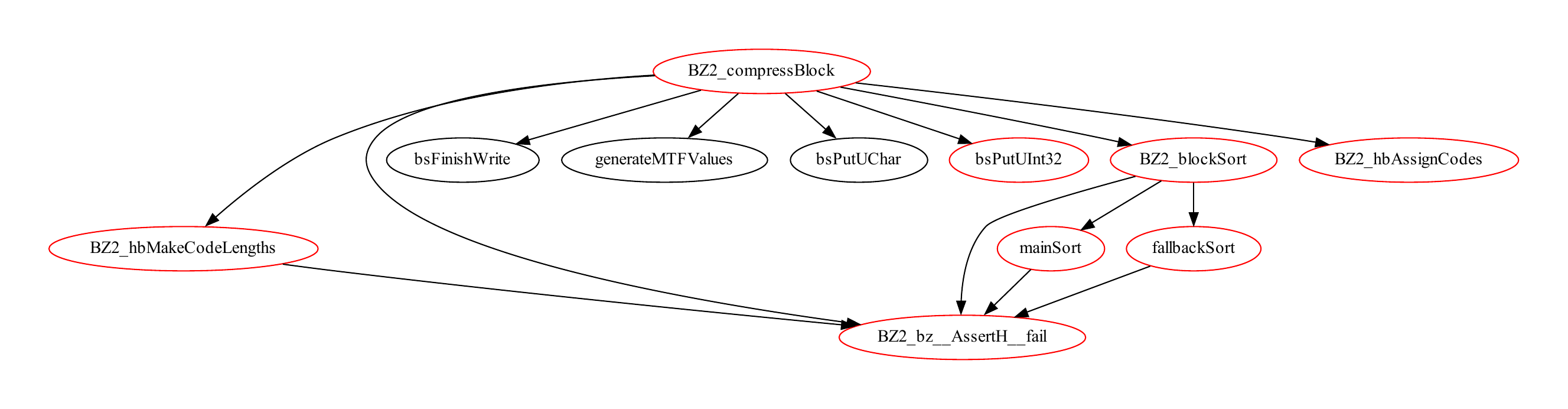}
			\label{fig:motivating_example1}
		\end{minipage}%
	}%
	\hfill
	\subfigure[FCG of BZ2\_compressBlock in Precomp (compiled by GCC using O2)]{
		\begin{minipage}[t]{0.85\linewidth}
			\centering
			\includegraphics[width=0.65\textwidth]{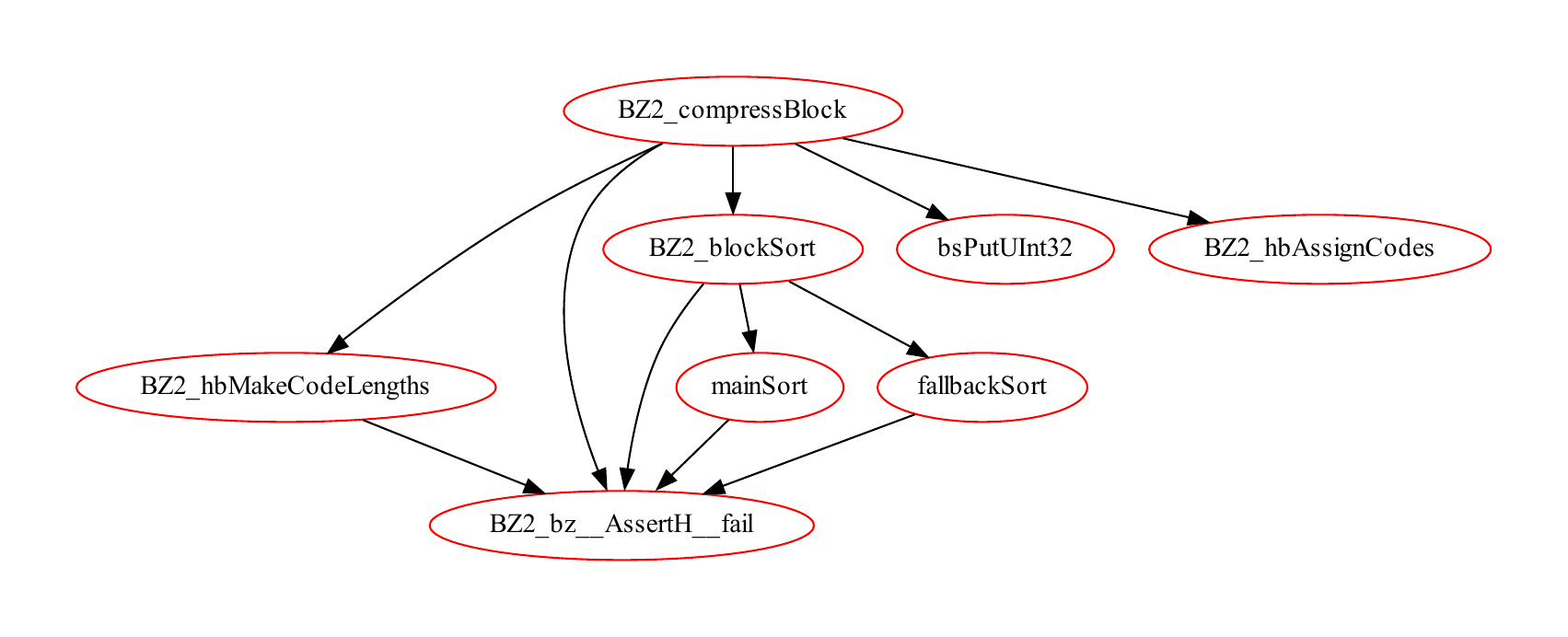}
			\label{fig:motivating_example2}
		\end{minipage}%
	}
	\vspace{-7pt}
	\caption{A motivating example for binary decomposition under compilation variance}
	\label{fig:motivation_example}
	\vspace{-10pt}
\end{figure*}

However, as binaries are compiled using diverse compilers, optimizations, and architectures, their function call relationships are also experiencing a great change. Figure~\ref{fig:motivation_example} shows a TPL reuse example between two compression software: \textit{Bzip2} and \textit{Precomp}. \textit{Precomp} reused the function \textit{BZ2\_compressBlock} from \textit{Bzip2}, and also reused all its callee functions. Even though \textit{Bzip2} and \textit{Precomp} are compiled with slightly different optimizations (O1 and O2), their FCGs have a great difference. When we use ModX~\cite{yang2022modx} to decompose these two FCGs, ModX classifies every function into separate communities, which leaves three functions (marked in grey) in \textit{Bzip2} that cannot find a match.

This difference can be mainly attributed to different function inlining decisions. Function inlining is a compiler optimization that replaces a function call with the insertion of the callee function body. When compiling \textit{Precomp} with GCC and O2, \textit{BZ2\_compressBlock} has inlined three of its callee functions (marked in grey in Figure~\ref{fig:motivating_example1}), making the FCG of \textit{Bzip2} has a different structure with the FCG of \textit{Precomp}. Moreover, the function \textit{BZ2\_bzCompress} in \textit{Precomp} also has different semantics when compared with the function \textit{BZ2\_bzCompress} in \textit{Bzip2}, as the function \textit{BZ2\_bzCompress} in \textit{Precomp} also contains the semantics from the inlined functions.

As the FCGs have changed when using different compilation settings, it raises a question of what is the best decomposition that can decompose these two FCGs to obtain semantically equivalent function modules. To our best knowledge, none of the existing works have constructed the oracle of best decomposition, nor have they systematically evaluated the existing binary decomposition works. 

In this work, we propose a method to identify the optimal decomposition. We first propose a labeling method to construct FCG mappings for a dataset compiled by 17 compilers, using 6 optimizations to 4 architectures. 
We then build the oracle by identifying stable boundary functions—those never inlined across any compilation setting—and generating Minimal Equivalent Function Regions (MEFRs) as ground-truth communities. Based on the oracle, we propose several evaluation metrics for anchor-based and clustering-based methods. Finally, using these evaluation metrics, we evaluate existing decomposition methods under compilation variance, revealing they suffer from either under-aggregation failure or over-aggregation failure.


Our main contributions are listed as follows:

\begin{enumerate}
	\item \textbf{Oracle Construction}:  We propose the first systematic methodology to build ground-truth binary decomposition oracles across compilation variants, addressing a fundamental gap in the field.
	
	\item \textbf{Labeled Dataset}: We propose a labeling method to construct FCG mappings and oracles for a dataset compiled by 17 compilers, using 6 optimizations to 4 architectures.

	\item \textbf{Empirical Evaluation}: Using our oracle, we reveal critical limitations of existing methods: they suffer from either under-aggregation failure or over-aggregation failure when compared to the oracle.
	
	\item \textbf{Public Artifacts}: We release the oracle construction toolchain, labeled dataset, and evaluation framework to support future research\footnote{\url{https://github.com/island255/Binary-Decomposition-Under-FCG-Variance}}.
	
\end{enumerate}

\section{Related Work}

As illustrated in the Section~\ref{sec:introduction}, the binary decomposition works under compilation variance face the challenge of FCG variance. As the FCG structure is mainly influenced by different inlining decisions, we will introduce the function inlining, binary decomposition works, and function inlining related binary code analysis works in this section.

\subsection{Function Inlining}

\label{sec:function_inlining}
Function inlining~\cite{Function_inlining} is a technique that replaces a function call with the body of the callee function. It can reduce the overhead of function calls and enable further compilation optimizations~\cite{DBLP:conf/cgo/ProkopecDLW19}. By inlining the callee function, the intra-procedural optimizations can be applied to functions generated by inlining, making the choice of optimizations more comprehensive.

However, function inlining may also bring side effects, such as increasing the binary code size and the compilation time. Therefore, it is a trade-off to decide whether to inline or not. Previous researches have proposed various inlining strategies~\cite{DBLP:journals/tse/DavidsonH92, DBLP:conf/lfp/DeanC94, hubicka2004gcc, DBLP:conf/lcpc/ZhaoA03, DBLP:conf/cc/CooperHW08, andersson2009evaluation, DBLP:conf/pldi/HwuC89, durand2018partial} from different perspectives. In this paper, we will introduce the inlining strategies used in GCC~\cite{gcc} and LLVM~\cite{LLVM}, two of the most popular compilers.

GCC and LLVM have a similar workflow for function inlining, which consists of two steps: inlining decision and inlining conduction. To make the inlining decision, they consider three types of inlining candidates: user-forced, user-suggested, and normal~\cite{LLVM_IPO1, GCC_inline}. User-forced inlining, such as ``$always\_inline$'', will be inlined as long as the call site does not violate some basic requirements. User-suggested inlining is marked by the keyword ``$inline$'' in the function definition. These functions may be inlined depending on the compiler's strategy. Normal inlining candidates are determined by the compiler's own heuristics.

GCC uses a priority-based inlining algorithm to decide which call sites to inline. It first estimates the inlining benefit of each call site and assigns a priority to it. The inlining cost also increases with each inlining. The inlining process stops when the total cost reaches a predefined threshold. Besides the general inlining strategy, GCC also has some special settings for small functions and functions that are called only once. For example, when compiling with the option of ``-O1", GCC applies ``-finline-functions-called-once'' to inline all static functions that are called only once, regardless of whether they are marked with ``inline'' or not. These functions will not be compiled as separate functions in the binary~\cite{GCC_inline}.

LLVM uses a prediction-based inlining method to decide which functions to inline. It first classifies functions into hot and cold ones based on their call frequencies. Then it calculates the inlining benefit and cost for each call site~\cite{InlineCost.cpp}. If the benefit is higher than the cost, the call site will be inlined~\cite{LLVM_inline}. Clang is the C/C++ frontend of the LLVM, and we will use Clang in the paper to represent the C/C++ toolchain of the LLVM.

\subsection{Binary Decomposition}

Binary decomposition serves as an important technique in the TPL detection. According to the language of TPL, TPL detection works can be classified into C/C++ TPL detection works~\cite{li2023libam, xu2021interpretation, dong2024libvdiff, tang2022libdb, zhu2022bbdetector, yang2022modx, sun2023moddiff, karande2018bcd, guo2023searching, tang2020libdx} and Java TPL detection works~\cite{backes2016reliable, li2017libd, ma2016libradar, zhan2021atvhunter, zhan2020automated, zhang2019libid, zhang2018detecting}. In this work, we will focus on binary decomposition works in C/C++ TPL detection, where the diverse compilation options and architectures exhibit greater challenges.
Currently, the fine-grained binary decomposition works can be classified into two classes: anchor-based methods and clustering-based methods.


Anchor-based methods~\cite{li2023libam, xu2021interpretation, dong2024libvdiff, tang2022libdb} conduct binary decomposition by first identifying some function pairs with high similarities as anchor nodes, and then comparing the neighbor nodes of anchor nodes to search for more matched functions. IRSD~\cite{xu2021interpretation} first identifies the functions that have the same instructions or have identical library calls as anchor functions, then compares functions that have the most anchor functions as neighbor nodes. LibDB~\cite{tang2022libdb} first identifies similar function pairs by selecting functions with a similarity larger than 80\%, then uses the FCG architecture to examine the matched functions. LibAM~\cite{li2023libam} first identifies the highly similar function pairs as anchor nodes, then extends anchor nodes to construct a function area, and finally compares the function area to identify reused TPLs. Though anchor-based methods use different strategies to detect reused TPLs, they all rely on the local FCG structure of anchor nodes to match more functions.


Clustering-based methods~\cite{yang2022modx, sun2023moddiff, karande2018bcd, guo2023searching} first cluster the binary functions into modules, and then compare these modules to identify reused TPLs. BCD~\cite{karande2018bcd} uses three properties, including code locality, data references, and function calls, to construct a graph for clustering. Then BCD uses Newman’s generalized community detection algorithm~\cite{newman2004fast} to group binary functions to generate modules in the binary. 
ModX~\cite{yang2022modx} first defines a module quality score to assess the coherence of the function clusters, and then it starts to group individual functions to form modules while maximizing the overall module quality score. 
BMVul~\cite{guo2023searching} introduces directed binary modularization (DBM), an overlapping community detection algorithm for binary function clustering. 
Though clustering-based methods use different features to construct their clustering graph, the structure of FCG is usually the most basic structure where the clustering starts.

\subsection{Binary Code Analysis Under Function Inlining}


Existing works~\cite{bingo, asm2vec, jia20231, jia2022comparing, jia2024cross, lin2024reifunc, sha2025optrans, qiu2015library, qiu2015using, ahmed2021learning, binosi2023bino, lin2023fsmell, dall2022highliner} have taken preliminary research of binary code analysis under function inlining. The first binary function similarity detection work that takes function inlining into consideration is Bingo~\cite{bingo}. Bingo summarizes several patterns that a caller function should inline its callee functions and conducts inlining to simulate the functions generated by function inlining. Asm2Vec~\cite{asm2vec} also uses Bingo's strategies to tackle function inlining. However, their simulation inlining strategies are proven inaccurate~\cite{jia20231}, thus cannot tackle the challenges that function inlining brings. 

Jia~\cite{jia20231} has conducted the first systematic empirical study to investigate the effect of function inlining on binary function similarity analysis. They evaluated four works on the dataset with inlining, and results show that most works suffer a 30\%-40\% performance loss when detecting the functions with inlining. To tackle the challenges that function inlining brings, they propose two methods named O2NMatcher~\cite{jia2022comparing} and CI-Detector~\cite{jia2024cross}, for binary2source and binary2binary function similarity detection. 

Moreover, there are other works that aim to detect inlined functions. ReIFunc~\cite{lin2024reifunc} identifies inlined functions by detecting repeated code fragments in binaries. A lot of works~\cite{qiu2015library, qiu2015using, ahmed2021learning, binosi2023bino} aim to identify inlined library functions. Besides, OpTrans~\cite{sha2025optrans} uses a binary rewriting technique to re-optimize binaries into similar inlining decisions.

However, the above works mostly focus on the function-level binary similarity analysis. None of these works has systematically explored the oracle for binary decomposition works, nor have they evaluated binary decomposition works under various compilation settings. In this work, we will conduct the first study to investigate these issues. Our work will provide a new sight for the binary decomposition works under compilation variance from the program-level view.

\section{Oracle Construction}

\subsection{Overview}
\label{sec:oracle_overview}

\begin{figure}[htbp]
	\centering
	\includegraphics[width=1\textwidth]{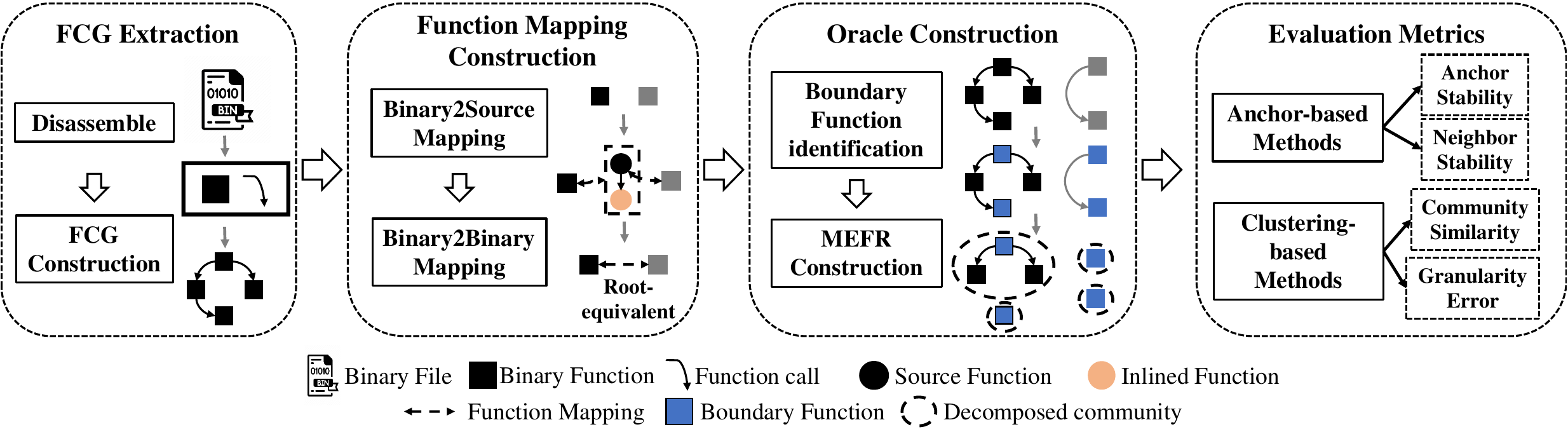}
	\vspace{-20pt}
	\caption{The overview of oracle construction}
	\label{fig:overview}
	\vspace{-10pt}
\end{figure}

Existing binary decomposition methods evaluate their effectiveness through indirect metrics (e.g., matching accuracy) rather than ground-truth decomposition quality. This gap severely hinders the community's ability to diagnose whether failures stem from algorithmic flaws or fundamental issues with FCG instability. To address this, we present the first systematic methodology to construct a \textbf{compilation-aware oracle} for binary decomposition. Our oracle is built upon the concept of \textbf{Minimal Equivalent Function Regions (MEFRs)}—the smallest semantically equivalent function communities across compilation variants. Figure~\ref{fig:overview} presents the overview of oracle construction.

As shown in Figure~\ref{fig:overview}, the construction and evaluation framework comprises four key steps: (1) extracting the FCG from binaries compiled by various compilations (\S\ref{sec:dataset}), (2) building cross-compilation function mappings (\S\ref{sec:function_mapping}), (3) identifying stable boundary functions and generating MEFRs as ground-truth communities (\S\ref{sec:Oracle_Construction}), and (4) proposing oracle-based metrics to rigorously evaluate both anchor-based and clustering-based decomposition methods (\S\ref{sec:metrics}). This comprehensive pipeline not only provides a reliable ground truth but also enables quantitative diagnosis of limitations in existing decomposition works.

\subsection{Dataset and FCG Extraction}
\label{sec:dataset}

We use the Binkit dataset~\cite{Binkit} as our research foundation. 
As shown in Table~\ref{tab:dataset}, it includes 51 GNU projects, including third-party libraries such as gsl, nettle, and libmicrohttpd covering scientific computing, cryptography, and network protocols. Then Binkit compiles them using 17 compilers, 6 optimizations, to 4 architectures, totaling 408 configurations, as shown in Table~\ref{tab:dataset_setting}. 
This generates 100,768 binaries comprising 43,312,875 functions and 187,959,002 calls. 
Note that the gsl package misses 8 binaries with \texttt{Ofast} due to compiler bugs~\cite{BinkitGithub}, making the number of binaries not an exact multiple of the compilation configurations.

\begin{table}[h]
	\caption{Projects and Versions in the Dataset}
	\vspace{-8pt}
	\centering
	\begin{tabular}{c|c|c|c|c|c|c|c}
		\hline
		Project & Version & Project & Version & Project & Version & Project & Version \\ \hline
		a2ps      & 4.14    & binutils      & 2.3     & bool          & 0.2.2   & ccd2cue       & 0.5     \\ \hline
		cflow     & 1.5     & coreutils     & 8.29    & cpio          & 2.12    & cppi          & 1.18    \\ \hline
		dap       & 3.1     & datamash      & 1.3     & direvent      & 5.1     & enscript      & 1.6.6   \\ \hline
		findutils & 4.6.0   & gawk          & 4.2.1   & gcal          & 4.1     & gdbm          & 1.15    \\ \hline
		glpk      & 4.65    & gmp           & 6.1.2   & gnu-pw-mgr    & 2.3.1   & gnudos        & 1.11.4  \\ \hline
		grep      & 3.1     & gsasl         & 1.8.0   & gsl           & 2.5     & gss           & 1.0.3   \\ \hline
		gzip      & 1.9     & hello         & 2.1     & inetutils     & 1.9.4   & libiconv      & 1.15    \\ \hline
		libidn    & 2.0.5   & libmicrohttpd & 0.9.59  & libtasn1      & 4.13    & libtool       & 2.4.6   \\ \hline
		libunistring & 0.9.10 & lightning    & 2.1.2   & macchanger    & 1.6.0   & nettle        & 3.4.1   \\ \hline
		osip      & 5.0.0   & patch         & 2.7.6   & plotutils     & 2.6     & readline      & 7       \\ \hline
		recutils  & 1.7     & sed           & 4.5     & sharutils     & 4.15.2  & spell         & 1.1     \\ \hline
		tar       & 1.3     & texinfo       & 6.5     & time          & 1.9     & units         & 2.16    \\ \hline
		wdiff     & 1.2.2   & which         & 2.21    & xorriso       & 1.4.8   &               &         \\ \hline
	\end{tabular}
	\label{tab:dataset}
\end{table}

\begin{table}[h]
	\caption{The compilation settings in the dataset}
	\vspace{-8pt}
	\centering
	\begin{tabular}{cc}
		\hline
		Compilation Settings    & Dataset                                                                                                                                                                                                                                      \\ \hline
		Compilers               & \begin{tabular}[c]{@{}c@{}}gcc-4.9.4, gcc-5.5.0, gcc-6.5.0, gcc-7.3.0, gcc-8.2.0, gcc-9.4.0, \\ gcc-10.3.0, gcc-11.2.0, clang-4.0, clang-5.0, clang-6.0, \\ clang-7.0, clang-8.0, clang-9.0, clang-10.0, clang-11.0, clang-12.0\end{tabular} \\ \hline
		Optimizations           & O0, O1, O2, O 3, Os, Ofast                                                                                                                                                                                                                   \\ \hline
		Architectures           & X86-32, X86-64, ARM-32, ARM-64                                                                                                                                                                                                               \\ \hline
	\end{tabular}
	\label{tab:dataset_setting}
\end{table}

This dataset is widely used in many binary code analysis works~\cite{gu2025uniasm, horimoto2024approach, cohen2025experimental, du2025degnn, li2025fus, collyer2024know, khan2024compiler, du2025gbsim, jiang2022ifattn, see2025enhancing, wang2024binenhance, du2023review, zeng2025cross, yang2024binary}. To our knowledge, compared with existing state-of-the-art datasets~\cite{saul2024function, NEURIPS2024_6bbefc73, 280046, 10.1145/3533767.3534367}, this dataset is the most comprehensive compilation-variation dataset available.

We extract FCGs using IDA Pro~\cite{IDAPro}. Then we represent functions as nodes and calls as edges. Since multiple calls may exist between two functions, we use NetworkX's MultiDiGraph~\cite{multidigraph, networkx} to preserve all edge information.

\subsection{Cross-Compilation Function Mapping Construction}
\label{sec:function_mapping}

Accurate function mappings across compilations are fundamental to Oracle construction. We first establish binary2source mappings using debug symbols, then derive binary2binary mappings.

\textbf{Binary2source function mapping.} The dataset is compiled with \texttt{-g} to produce a \texttt{.debug\_line} section containing binary-address-to-source-line mappings. Note that the debug information is only available at the labeling phase, which will be stripped before conducting TPL detection.  Following Jia et al.~\cite{jia20231}, we link each binary address to its corresponding binary function and each source line to its corresponding source function. By extending the binary-address-to-source-line mappings to binary-function-to-source-function mappings, we construct the binary2source function mappings. Binary functions mapping to \textit{multiple} source functions indicate inlining, while binary functions mapping to one source function represent without inlining.

\textbf{Binary2binary function mapping.} By intersecting the binary2source mappings of two compilation variants, we establish precise binary2binary function correspondences. In general, we classify the binary2binary function mappings into three classes: identical mapping, root-equivalent mapping, and relevant mapping.

\begin{figure*}[b]
	\centering
	\subfigure[Identical]{
		\begin{minipage}[t]{0.23\linewidth}
			\centering
			\includegraphics[width=0.85\textwidth]{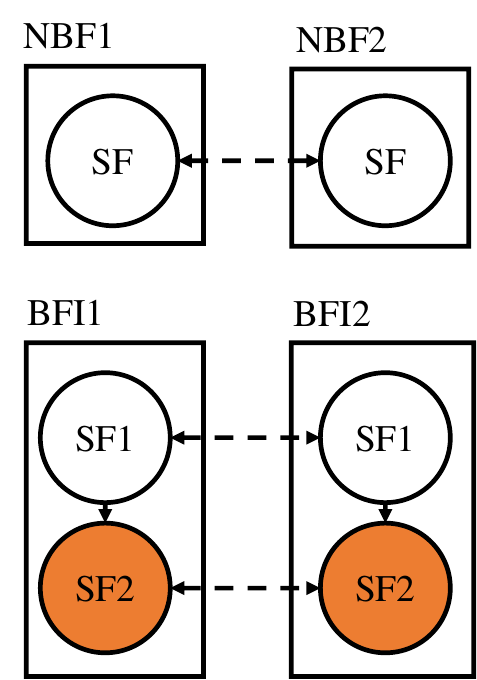}
			\label{fig:identical}
		\end{minipage}%
	}%
	\subfigure[Root-equivalent]{
		\begin{minipage}[t]{0.23\linewidth}
			\centering
			\includegraphics[width=0.85\textwidth]{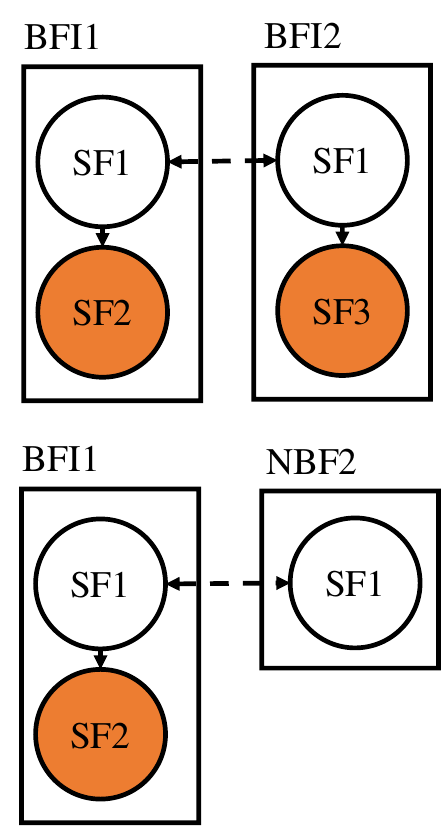}
			\label{fig:root-equivalent}
		\end{minipage}%
	}
	\subfigure[Relevant]{
		\begin{minipage}[t]{0.46\linewidth}
			\centering
			\includegraphics[width=0.85\textwidth]{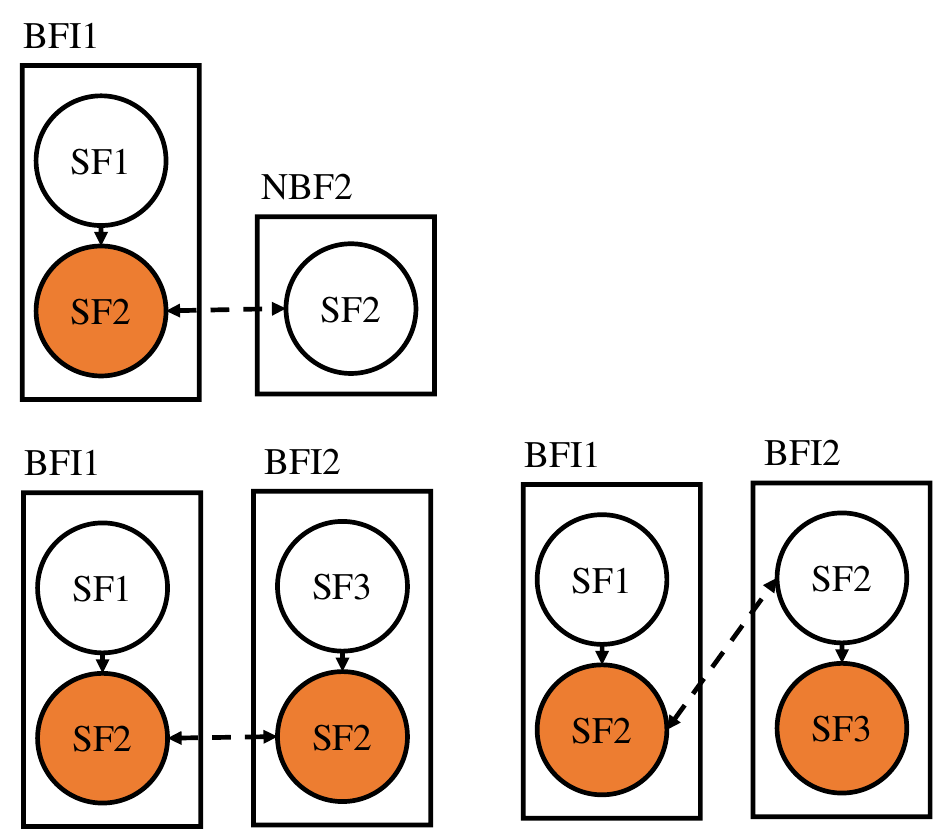}
			\label{fig:relevant}
		\end{minipage}%
	}
	\vspace{-10pt}
	\caption{ The binary2binary function mapping types}
	\label{fig:function_mapping_type}
\end{figure*}

Figure~\ref{fig:function_mapping_type} illustrates the three types of function mappings. Rectangles represent binary functions, and circles represent source functions. The circles inside the rectangles indicate the binary2source function mappings, meaning the binary function represented by the rectangle is compiled from the source functions represented by the circles inside it. For brevity, we use SF to represent the source function, NBF to represent the normal binary function (i.e., compiled without inlining), and BFI for the binary function generated by function inlining. The inlined source functions are marked in orange.

\begin{enumerate}
	
	\item \textbf{Identical Mappings (Figure~\ref{fig:identical}):}  When two binary functions map to the same set of source functions, we establish an identical mapping between them. Specifically, when two binary functions are both NBFs, if two binary functions compiled by different compilation settings map to the same source function, we will build an identical mapping between them. When two binary functions are both BFIs, and they also map to the same set of source functions, they will also be recognized as an identical mapping. For example, in Figure~\ref{fig:motivation_example}, the binary function \textit{fallbackSort} has not inlined source functions in \textit{Bzip} and \textit{Precomp}, which means that binary function \textit{fallbackSort} in \textit{Bzip} and \textit{Precomp} both map to the source function \textit{fallbackSort}, thus this function pair will have an identical mapping. 
	
	\item \textbf{Root-equivalent Mappings (Figure~\ref{fig:root-equivalent}):} When at least one binary function is a BFI, we construct the mapping based on their commonly mapped source functions. Specifically, we identify the original source function (OSF) of the BFI, which has the same name as the BFI. We then identify the source functions that these two binary functions both map to. If these two binary functions have the same OSF but their inlined source functions are not the same, we establish a root-equivalent mapping between them. For example, in Figure~\ref{fig:motivation_example}, the binary function \textit{BZ2\_compressBlock} in \textit{Bzip} and \textit{Precomp} is compiled from the same source function but has inlined different source functions, so they have a root-equivalent mapping.
	
	\item \textbf{Relevant Mappings (Figure~\ref{fig:relevant}):} If these two binary functions both map to some source functions but their OSFs are different, we establish a relevant mapping between them. For example, in Figure~\ref{fig:motivation_example}, the binary function \textit{BZ2\_compressBlock} in \textit{Precomp} has inlined the source function \textit{bsFinishWrite}, while the source function \textit{bsFinishWrite} is compiled to the binary function \textit{bsFinishWrite} in \textit{Bzip}. Therefore, the binary function \textit{bsFinishWrite} in \textit{Bzip} and the binary function \textit{BZ2\_compressBlock} in \textit{Precomp} have a relevant mapping.
	
\end{enumerate}

This classification enables us to track how a single source function manifests across different compilation settings, particularly under inlining transformations.

\subsection{Oracle Construction}
\label{sec:Oracle_Construction}

We formally define the MEFR as the oracle that we will construct:

\begin{definition}[MEFR]
	Given two binary FCGs $G_1$ and $G_2$ compiled from the same source code with different settings, a \textbf{Minimal Equivalent Function Region (MEFR)} is a subgraph $S_1 \subseteq G_1$ and its corresponding subgraph $S_2 \subseteq G_2$ such that: (1) $S_1$ and $S_2$ are semantically equivalent (i.e., they originate from identical source-level functions), and (2) no proper subgraph of $S_1$ or $S_2$ satisfies condition (1).
\end{definition}

\textbf{Boundary Function Identification.} The core insight behind our Oracle is that \textit{binary functions which have identical mapping or root-equivalent mapping and whose OSFs do not have relevant mappings with other binary functions can serve as stable anchors} across compilation variants. This insight originates from our function mapping analysis, where we derive two key observations:

\begin{enumerate}[label=(\alph*)]
	\item If binary functions $c_1$ and $c_2$ originate from the same OSF $c$, and $c$ is not inlined into \textit{other binary functions} (i.e., never-inlined source function $c$), then the subgraphs rooted at $c_1$ and $c_2$ are semantically equivalent. \label{obs:root}
	
	\item Similarly, if $a_1$ and $a_2$ originate from never-inlined source function $a$, then the FCG regions bounded by $a_1$-$c_1$ (the subgraph starts from  $a_1$ to $c_1$, which contains $a_1$ but does not contain $c_1$) and $a_2$-$c_2$ are semantically equivalent. \label{obs:region}
\end{enumerate}

We call  $a_1$, $a_2$, $c_1$, and $c_2$ as boundary functions.
Functions that are inlined in \textit{any} compilation settings cannot serve as boundary functions, as they either disappear or duplicate across multiple callers. Conversely, functions \textit{never inlined in any settings} provide stable boundaries for MEFR generation.

We identify these stable boundary functions by traversing all binary functions that have identical mapping or root-equivalent mapping, and their OSF does not have relevant mappings with other binary functions. The function regions surrounded by these boundary functions will form the MEFR that can serve as our oracle.

\textbf{MEFR Construction.}  
Algorithm~\ref{alg:community_partition} generates MEFRs from the identified set of boundary functions $P$. The algorithm employs a breadth-first search (BFS) strategy: each boundary function serves as the entry point of a region, and the algorithm traverses its successors level-by-level until encountering another boundary function or a terminal node, thereby preventing cross-region contamination.

Formally, the algorithm maintains a queue $Q$ initialized with all boundary functions. For each dequeued node $v$, it constructs a region $m$ starting from $v$. The inner loop processes a frontier set $S$ (initially $v$'s direct successors). Each node $s \in S$ is examined: if $s$ is not a boundary function and has not been included in $m$, it is added to the current region, and \textit{all its successors are enqueued into $S$} for subsequent processing. This BFS-style expansion ensures that \textit{all nodes reachable from $v$ via non-boundary paths} are captured, while termination upon reaching any boundary function guarantees semantic isolation between regions.

The examination of  $s \notin m$ ensures that it avoids the \textit{recursive duplication} problem, where deeply nested call chains might cause stack overhead or repeated node visits. By explicitly managing the frontier set $S$, the algorithm achieves deterministic O(|V|+|E|) time complexity per region and ensures that the generated MEFRs are stable across different execution orders.

The resulting MEFR set $\textsc{M}$ forms a partition of the function call graph: regions are mutually exclusive (no overlapping nodes) and collectively exhaustive (every node belongs to exactly one MEFR). Each region satisfies the two core properties of the MEFR definition: \textit{semantic equivalence} across compilation variants (by construction from stable boundary functions) and \textit{minimality} (no proper subgraph can preserve semantic equivalence). This rigorous construction provides the foundation for the oracle-based evaluation metrics in \S\ref{sec:metrics}.

\newcommand{\INPUT}{\item[\textbf{Input:}]}
\newcommand{\OUTPUT}{\item[\textbf{Output:}]}

\renewcommand{\algorithmiccomment}[1]{\hfill\makebox[9.5cm][l]{\textit{// #1}}}

\begin{algorithm}[!h]
	\caption{MEFR Construction}
	\label{alg:community_partition}
	\begin{algorithmic}[1]
		\INPUT function call graph $G = (V, E)$, set of boundary functions $P \subseteq V$
		\OUTPUT MEFR set $M$
		
		\STATE $M \gets \emptyset$
		\STATE $Q \gets P$ \COMMENT{Initialize queue with boundary functions}
		\WHILE{$Q$ is not empty}
		\STATE $v \gets \text{dequeue}(Q)$
		\STATE $m \gets \{v\}$ \COMMENT{Start from boundary function}
		\STATE $S \gets \text{successors of } v \text{ in } G$
		\WHILE{$S$ is not empty}
		\STATE $s \gets \text{dequeue}(S)$
		\IF{$s \notin P$  \AND $s \notin m$}  
		\STATE \COMMENT{Stop at boundary functions and avoid duplicates}
		\STATE $m \gets m \cup \{s\}$
		\STATE $S \gets S \cup \text{successors of } s$
		\ENDIF
		\ENDWHILE
		\STATE $M \gets M \cup \{m\}$ \COMMENT{Add generated MEFR}
		\ENDWHILE
		\RETURN $M$
	\end{algorithmic}
\end{algorithm}

\textbf{Oracle Validation.} To ensure our Oracle meets the MEFR definition, we randomly sample 100 FCG pairs and manually verify their generated regions. Results confirm that: (1) all MEFRs are semantically equivalent across pairs, and (2) no MEFR can be further subdivided while preserving semantic equivalence. This validation confirms that our Oracle provides a reliable ground truth for evaluating binary decomposition methods.

Unlike anchor-based methods that rely on heuristics to select anchors, or clustering-based methods that generate communities without semantic guarantees, our Oracle-derived MEFRs represent \textit{provably minimal and semantically equivalent} regions. This enables rigorous evaluation: any decomposition that splits an MEFR loses semantic completeness, while merging multiple MEFRs sacrifices detection granularity. Our Oracle thus serves as both an upper bound on achievable accuracy and a diagnostic tool to identify specific failure modes in existing works.

\subsection{Evaluation Metrics}
\label{sec:metrics}

We propose the first oracle-grounded metric suite that quantifies \emph{absolute} rather than relative decomposition quality.  
All scores are computed against the MEFR set $\mathcal{M}$ derived in \S\ref{sec:Oracle_Construction}, enabling us to pinpoint whether failures stem from algorithmic defects or fundamental information loss induced by inlining.

\subsubsection{Anchor-Based Methods}
\label{sec:metrics-anchor}

Anchor-based methods assume that (i) function names are stable anchors and (ii) neighbourhoods remain topologically consistent.  
We evaluate each assumption separately.

\textbf{Anchor Stability} ($S_{\text{anch}}$) measures how many \emph{true} anchors survive compilation variance:
\begin{equation}
	S_{\text{anch}} = \frac{|V_{1}^{\text{bf}} \cap V_{2}^{\text{bf}}|}{|V_{1}^{\text{bf}} \cup V_{2}^{\text{bf}}|},
	\label{eq:anch-stab}
\end{equation}
where $V_{i}^{\text{bf}}$ is the set of functions in FCG $G_i$, $V_{1}^{\text{bf}} \cap V_{2}^{\text{bf}}$ calculate the set of \emph{boundary functions} (identical or root-equivalent, never-inlined) in FCG $G_1$ and $G_2$, and $V_{1}^{\text{bf}} \cup V_{2}^{\text{bf}}$ calculate the set of all functions in FCG $G_1$ and $G_2$.

Low $S_{\text{anch}}$ directly flags that the method is attempting to anchor on functions that \emph{do not exist} in the second binary---a failure mode invisible to prior accuracy-based scores.

\textbf{Neighbour Stability} ($S_{\text{nb}}$) evaluates whether the local topology around a \emph{matched} anchor remains intact:
\begin{equation}
	S_{\text{nb}}(u,v) = \frac{|N(u) \cap N(v)|}{|N(u) \cup N(v)|}, \quad u,v\ \text{boundary functions},
	\label{eq:nb-pres}
\end{equation}
where $N(u)$ denotes the set of neighbor nodes of boundary functions $u$, $N(u) \cap N(v)$ represents the set of boundary functions in the neighbor nodes of boundary functions, and  $N(u) \cup N(v)$ represents the set of all functions in the neighbor nodes of boundary functions.

Low $S_{\text{nb}}(u,v)$ directly flags that the method is attempting to extend on functions that \emph{do not exist} in the second binary---a failure mode invisible to prior accuracy-based scores.

\subsubsection{Clustering-Based Methods}
\label{sec:metrics-cluster}

To rigorously evaluate clustering-based methods, we introduce an oracle-based scoring scheme that does not require internal algorithm knowledge. This approach allows us to systematically assess the performance of clustering tools against ground-truth mappings identified by our oracle.

\textbf{Step 1: Source-function set extraction.}
We begin by extracting the semantic footprint of each community produced by the clustering tool. Let
\begin{equation}
	\text{SF}(C)=\!\!\bigcup_{\text{node}\,\in\,C}\!\!\text{SF}(\text{node}),
	\label{eq:com-sf}
\end{equation}
where $\text{SF}(\text{node})$ represents the set of source-level functions mapped by the binary node (binary function). Thus, $\text{SF}(C)$ encapsulates the semantic content of community $C$.

\textbf{Step 2: Similarity Calculation.}
We define the similarity between two communities $C_1$ and $C_2$ based on their semantic footprints:
\begin{equation}
	\mathcal{S}(C_1, C_2) = \frac{|\text{SF}(C_1)\cap\text{SF}(C_2)|}{|\text{SF}(C_1)\cup\text{SF}(C_2)|}.
	\label{eq:similarity}
\end{equation}
This metric quantifies the overlap between the semantic contents of the two communities, providing a measure of how well the clustering aligns with the oracle-derived ground truth.

\textbf{Step 3: Nearest Community Identification.}
For each ground-truth Minimal Equivalent Function Region (MEFR) $M\in\mathcal{M}$, we identify the generated community that best recovers it:
\begin{equation}
	\mathcal{C}(M)=\arg\max_{C\,\in\,\mathcal{C}_{\text{gen}}}\mathcal{S}(M, C).
	\label{eq:coverage}
\end{equation}
This step ensures that we compare each ground-truth MEFR with the most similar community produced by the clustering method, allowing us to evaluate the method's ability to capture semantically equivalent regions.

\textbf{Step 4: Granularity Error Measurement.}
To detect over-aggregation or under-aggregation, we measure the granularity error as:
\begin{equation}
	\mathcal{G}(M)=\frac{|\text{SF}(C^{\star})|}{|\text{SF}(M)|},\quad C^{\star}=\arg\max_{C}\,\mathcal{S}(M, C).
	\label{eq:gran}
\end{equation}
Here, $\mathcal{G}(M)>1$ indicates that several MEFRs have been merged into one community (over-aggregation), while $\mathcal{G}(M)<1$ suggests that a single MEFR has been split (under-aggregation). This metric provides insight into the clustering method's tendency to either over-aggregate or over-segment the function mappings.

With this oracle, existing works could detect this fundamental trade-off, revealing the hidden defects that lead to poor performance. By systematically applying these metrics, we can expose critical limitations in clustering-based binary decomposition methods and guide the development of more robust algorithms.

\section{Evaluation Results}
\label{sec:evaluation}

To validate the effectiveness of our oracle and demonstrate its diagnostic power, we pose and answer three oracle-driven research questions:

\begin{itemize}
	\item \textbf{RQ1}: What ground-truth function mapping distributions does our oracle reveal across compilation settings?
	\item \textbf{RQ2}: How stable are the anchors used by existing anchor-based methods according to our oracle?
	\item \textbf{RQ3}: How does our oracle help reveal the failure modes of clustering-based methods?
\end{itemize}

We first validate our oracle by quantifying the compilation-induced mapping complexity it captures (RQ1). Then we demonstrate its diagnostic value by using it to systematically evaluate and expose critical limitations in both anchor-based and clustering-based methods (RQ2--RQ3).

\subsection{RQ1: What ground-truth function mapping distributions does our oracle reveal?}
\label{sec:rq1}

\begin{figure*}[h]
	\centering
	\subfigure[Cross-optimization]{
		\begin{minipage}[t]{1\linewidth}
			\centering
			\includegraphics[width=0.95\textwidth]{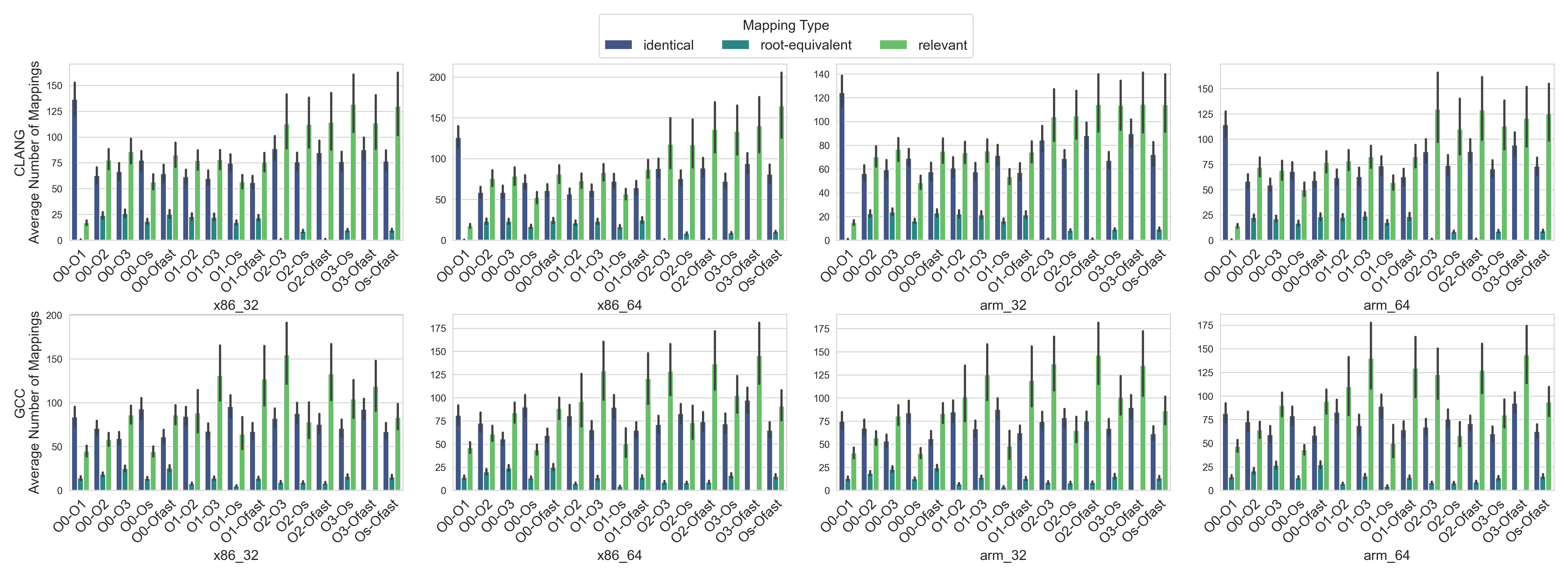}
			\label{fig:cross_opt}
		\end{minipage}%
	}%

	\centering
	\vspace{-5pt}
	\subfigure[Cross-compiler: top 10 compilation pairs with the highest root-equivalent mapping ratios]{
		\begin{minipage}[t]{1\linewidth}
			\centering
			\includegraphics[width=0.75\textwidth]{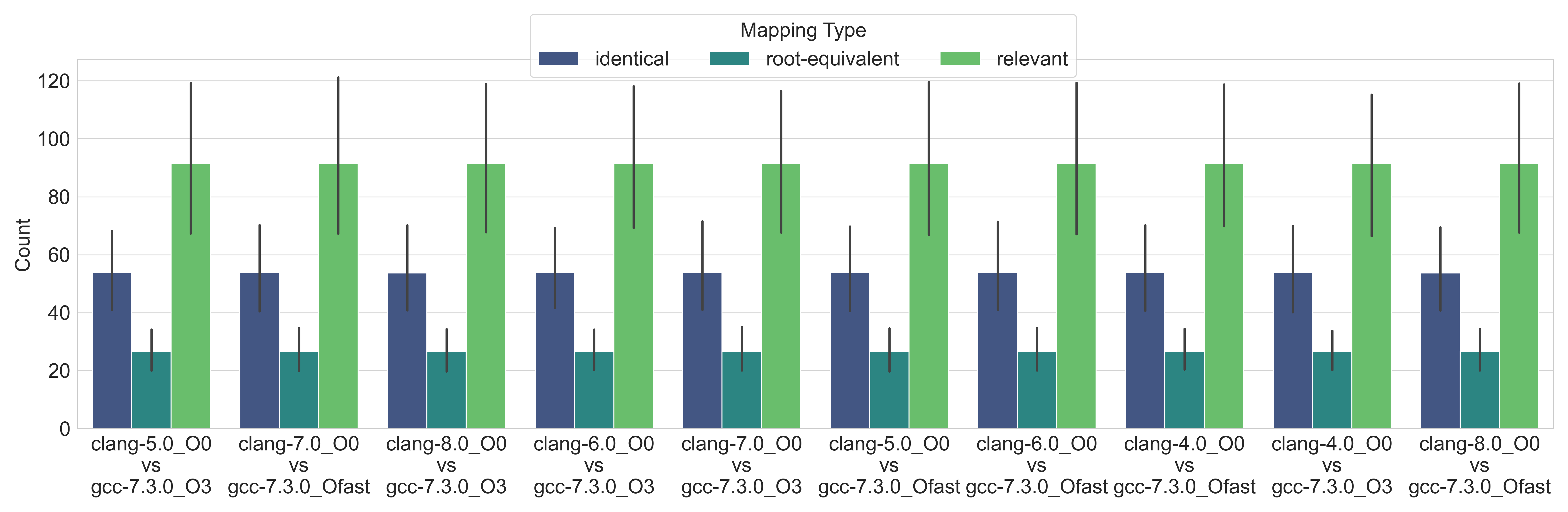}
			\label{fig:cross_compiler}
		\end{minipage}%
	}

	\centering
	\vspace{-5pt}
	\subfigure[Cross-architecture]{
		\begin{minipage}[t]{1\linewidth}
			\centering
			\includegraphics[width=0.95\textwidth]{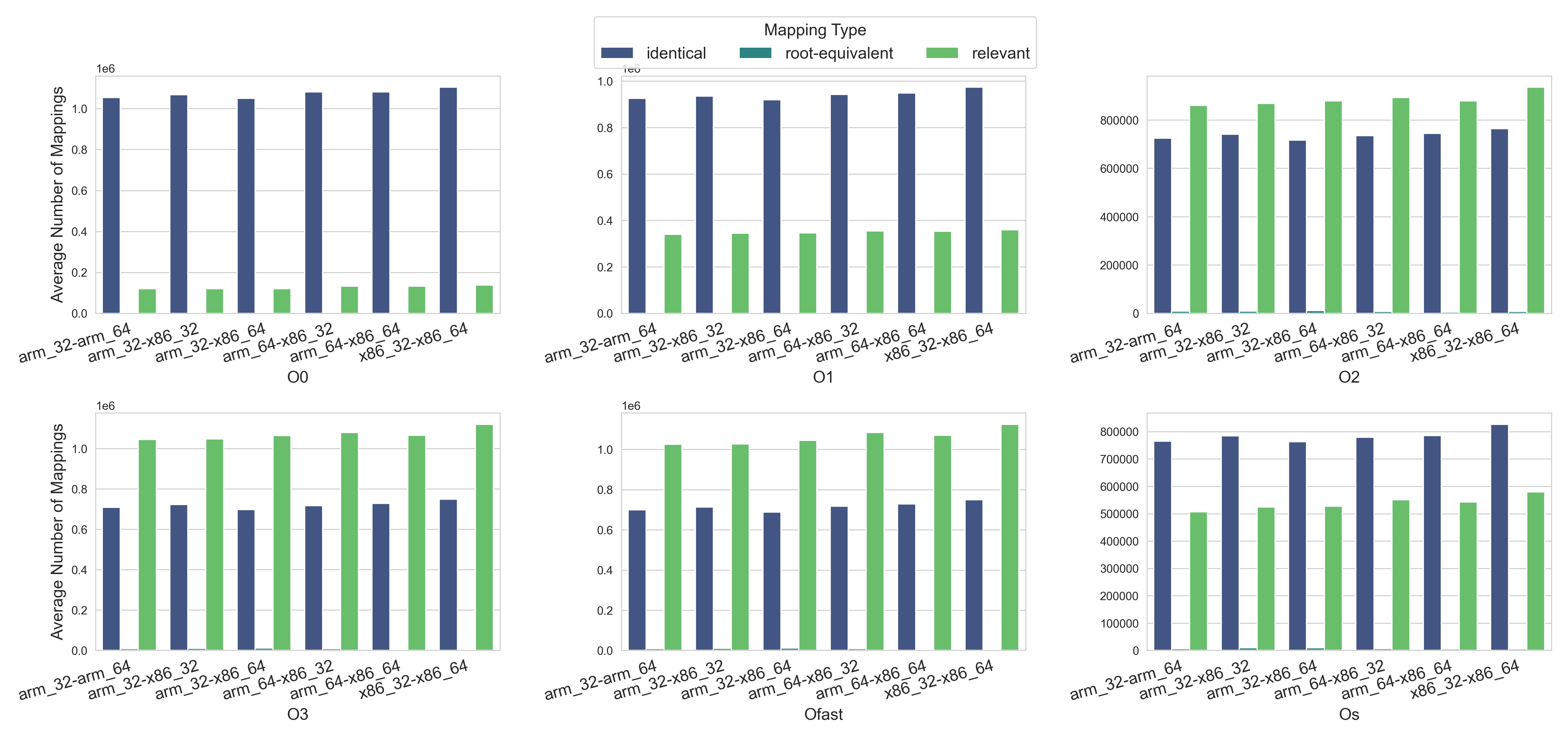}
			\label{fig:cross_arch}
		\end{minipage}%
		\vspace{-5pt}
	}	
	
	\vspace{-10pt}
	\caption{Oracle-derived mapping distributions in cross-compilation scenarios}
	\label{fig:cross_function_mapping}
	\vspace{-5pt}
	
\end{figure*}

Our oracle enables the precise quantification of how compilation settings fragment function mappings. We analyze the distribution of mapping types across various compiler-optimization pairs and architectures. The results are presented in Figure~\ref{fig:cross_function_mapping}.

\textbf{Escalating Fragmentation in Cross-Optimization (Figure~\ref{fig:cross_opt}).} The oracle reveals a stark gradient in mapping stability. When comparing O0 vs. O1, our ground truth shows \textit{identical mappings} dominate across all architectures, accounting for 75-150 mappings per binary, while root-equivalent mappings remain negligible (<25). However, as the optimization gap widens, the oracle captures severe fragmentation: O0 vs. Ofast comparisons show identical mappings \textit{plummeting} to 50-75, while root-equivalent mappings surge to about 25 and relevant mappings surge to 75-100. The oracle precisely quantifies this degradation—a 50\% loss of one-to-one function mapping when aggressive inlining is introduced.

\textbf{Amplified Complexity in Cross-Compiler (Figure~\ref{fig:cross_compiler}).} Figure~\ref{fig:cross_compiler} shows the top 10 compilation pairs with the highest root-equivalent mapping ratios.   Our oracle reveals that compiler family differences significantly amplify mapping complexity beyond optimization effects alone. Our observation finds that Clang vs. GCC comparisons, at the most challenging scenario, O0 (Clang) vs. Ofast (GCC), show the oracle identifying 15-25 mappings as identical, nearly half of the identical mappings, with relevant mappings becoming the dominant category. This ground-truth data confirms that divergent inlining heuristics between compiler families create lots of partial function overlaps that greatly challenge existing methods.

\textbf{Preserved Stability in Cross-Architecture (Figure~\ref{fig:cross_arch}).} The oracle reveals that different architectures \textit{preserve} mapping type distributions, validating our focus on compilation variance as the primary challenge. For x86\_32-x86\_64 comparisons, the number of identical mappings remains stable across optimizations. ARM comparisons show similar trends, though with slightly higher root-equivalent mappings due to architecture-specific intrinsics being inlined differently. The oracle confirms that the FCG structure is largely maintained across architectures, making compiler and optimization settings the dominant source of variance.


\textbf{Observation 1.1: Low-optimization comparisons preserve structure.} When comparing O0 vs. O1 in Clang, identical mappings dominate, confirming the oracle correctly captures minimal inlining.

\textbf{Observation 1.2: Aggressive optimization fragments functions severely.} Our oracle reveals that low vs. high optimization comparisons (e.g., O0 vs. O2) contain only 38\% identical mappings, with 50\% relevant and 12\% root-equivalent mappings. This ground-truth data quantifies the exact challenge decomposition methods face.

\textbf{Observation 1.3: Cross-compiler mappings are most complex.} The oracle identifies that Clang-vs-GCC comparisons produce root-equivalent mappings up to half of the identical mappings—the highest complexity level—due to divergent inlining strategies.

\textbf{Answer to RQ1:} The oracle provides the first quantitative ground truth revealing that compilation variance yields \textless40\% identical mappings in challenging scenarios, with cross-compiler settings producing root-equivalent mappings up to half of the identical mappings. It also confirms that function fragmentation is driven primarily by inlining heuristics rather than architecture, and captures fine-grained differences invisible to prior methods. This oracle-derived benchmark establishes the precise difficulty curve that decomposition methods must overcome.

\subsection{RQ2: How stable are anchors according to our oracle?}
\label{sec:rq2}

Our oracle defines boundary functions as stable anchors (functions never inlined in any setting). We use this ground truth to evaluate the anchor stability assumption that anchor-based methods rely upon.

\subsubsection{Cross-optimization Stability}

\begin{figure}[t]
	\centering
	\subfigure[Anchor stability - Clang]{
		\begin{minipage}[t]{0.48\linewidth}
			\centering
			\includegraphics[width=0.95\textwidth]{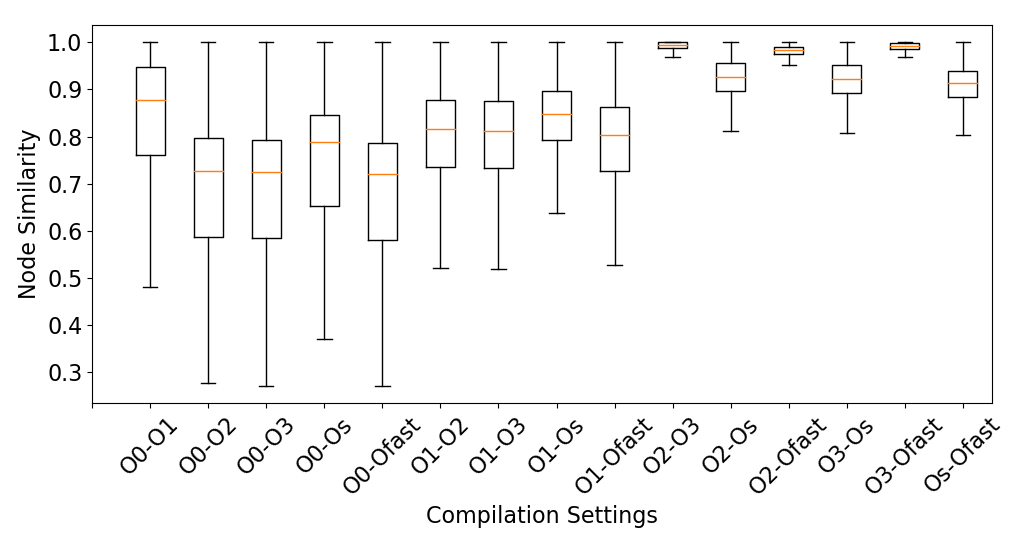}
			\label{fig:clang_node_sim}
			\vspace{-15pt}
		\end{minipage}%
	}
	\subfigure[Anchor stability - GCC]{
		\begin{minipage}[t]{0.48\linewidth}
			\centering
			\includegraphics[width=0.95\textwidth]{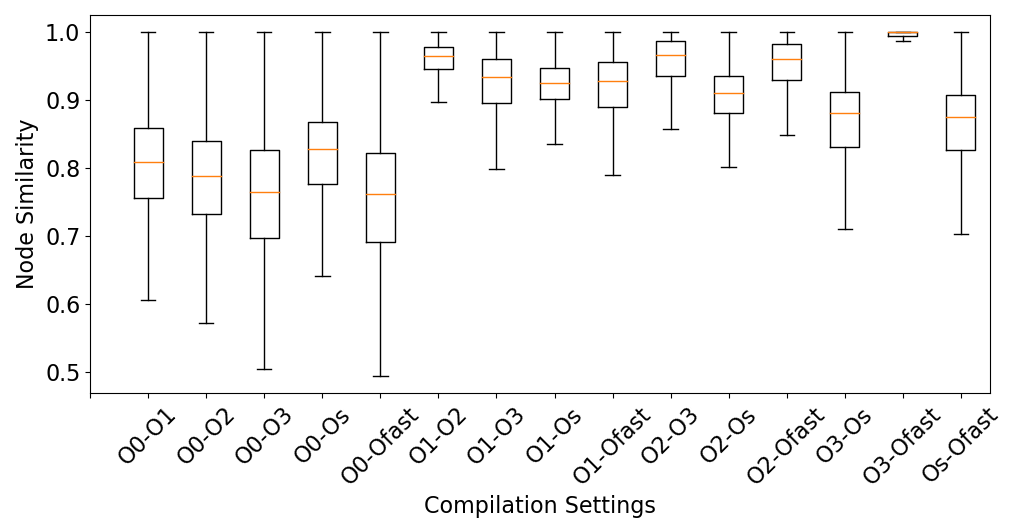}
			\label{fig:gcc_node_sim}
			\vspace{-15pt}
		\end{minipage}%
	}

	\subfigure[neighbor stability - Clang]{
		\begin{minipage}[t]{0.48\linewidth}
			\centering
			\includegraphics[width=0.95\textwidth]{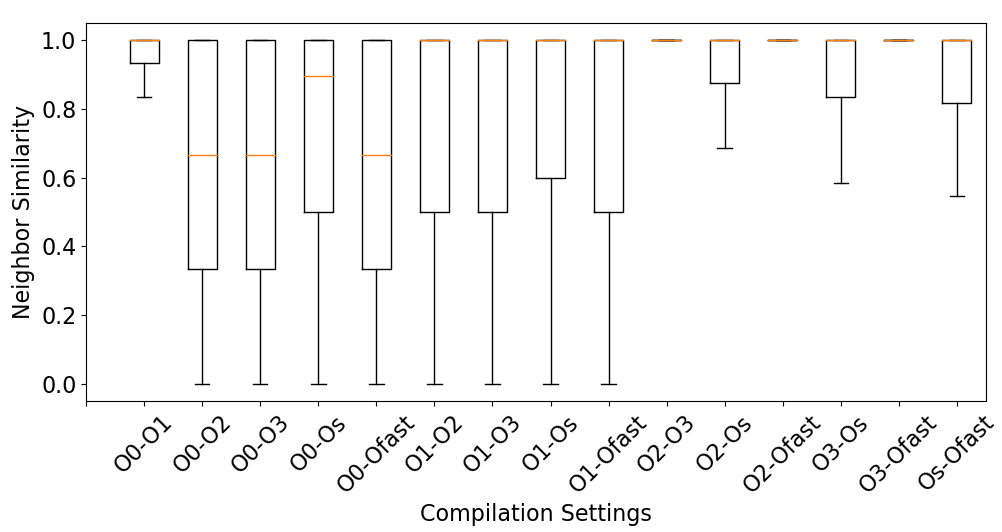}
			\label{fig:clang_neighbor_sim}
			\vspace{-15pt}
		\end{minipage}%
	}%
	\subfigure[neighbor stability - GCC]{
		\begin{minipage}[t]{0.48\linewidth}
			\centering
			\includegraphics[width=0.95\textwidth]{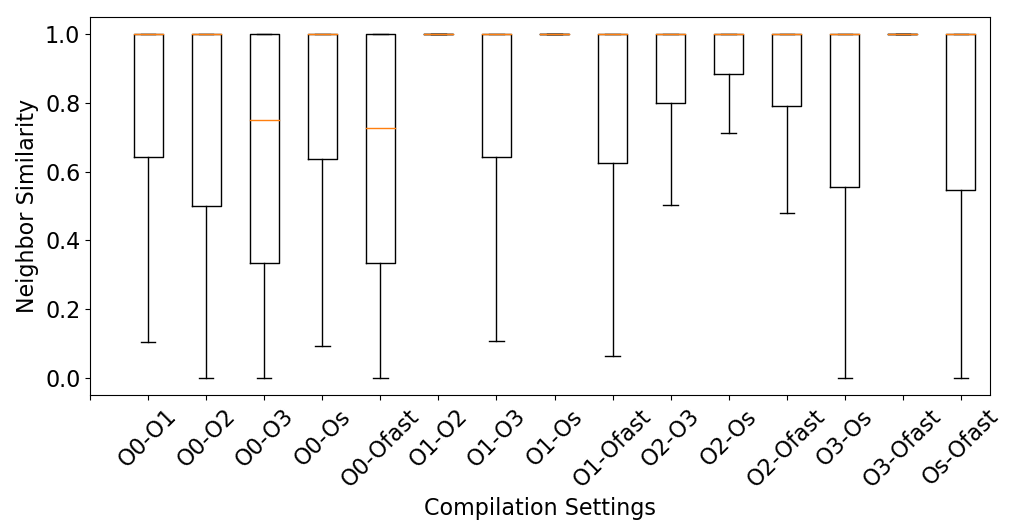}
			\label{fig:gcc_neighbor_sim}
			\vspace{-15pt}
		\end{minipage}%
	}%
	\vspace{-10pt}
	\caption{Oracle-measured anchor stability and neighbor stability in cross-optimization evaluation}
	\vspace{-10pt}
	\label{fig:node_neighbor}
\end{figure}

Figure~\ref{fig:node_neighbor} shows the oracle-measured anchor stability and neighbor stability in the cross-optimization evaluation of Clang and GCC. In Figure~\ref{fig:node_neighbor}, Figure~\ref{fig:clang_node_sim} shows the measured anchor stability in Clang, Figure~\ref{fig:gcc_node_sim} shows the measured anchor stability in GCC, Figure~\ref{fig:clang_neighbor_sim} shows the measured neighbor stability in Clang, and Figure~\ref{fig:gcc_neighbor_sim} shows the measured neighbor stability in GCC. We use anchor similarity to measure the anchor stability and the neighbor similarity for the neighbor stability. 

As shown in Figure~\ref{fig:clang_node_sim} and Figure~\ref{fig:gcc_node_sim}, when comparing low (O0, O1) vs. high optimizations(O2, O3, Os, Ofast), only 70-80\% of binary functions have their anchor functions. More critically, our oracle's semantic labels reveal that 25\% of these are root-equivalent mappings on average (as shown in Figure~\ref{fig:cross_opt}), meaning \textit{only 52.5-60\% are truly stable anchors}.

Figure~\ref{fig:clang_neighbor_sim} and Figure~\ref{fig:gcc_neighbor_sim} show the distribution of neighbor similarities when comparing different optimizations. As shown in the figure, FCGs compiled using adjacent optimizations share similar neighbor functions. However, when comparing low optimizations with high optimizations, such as O0 to O3, the neighbor similarities in average can be less than 70\%. Moreover, there are many anchor functions where they do not share any common neighbor functions, which may pose a great challenge to anchor-based methods, as there are no neighbor functions that can be further matched.

\subsubsection{Cross-compiler Stability}

In the cross-compiler evaluation, we select the top 10 lowest anchor similarities and neighbor similarities to reveal the anchor stability and neighbor stability of anchor-based methods. As our dataset includes 17 compilers and 6 optimizations, the top 10 results are selected from $17 * 6 * (17 * 6 - 1)  / 2  = 5199 $ pairs of cross-compiler detection cases. Table~\ref{tab:top10_lowest_node_sim} and Table~\ref{tab:top10_lowest_neighbor_sim} shows the results.

\begin{minipage}[h]{\textwidth}
	\vspace{7pt}
	\centering
	\begin{minipage}[]{0.48\textwidth}
		\centering
		\makeatletter\def\@captype{table}\makeatother\caption{Top 10 lowest anchor similarities in cross-compiler evaluation}
		\vspace{-5pt}
		\scalebox{0.85}{
			\begin{tabular}{c|c|c|c}
				\hline
				rank & compiler-opt 1 & compiler-opt 2 & anchor similarity \\ \hline
				1.   & gcc-11.2.0-Ofast      & clang-6.0-O0          & 0.6281     \\ \hline
				2.   & gcc-11.2.0-Ofast      & clang-7.0-O0          & 0.6282     \\ \hline
				3.   & gcc-11.2.0-Ofast      & clang-4.0-O0          & 0.6283     \\ \hline
				4.   & gcc-11.2.0-Ofast      & clang-5.0-O0          & 0.6285     \\ \hline
				5.   & gcc-11.2.0-O3         & clang-6.0-O0          & 0.6319     \\ \hline
				6.   & gcc-11.2.0-O3         & clang-7.0-O0          & 0.6320     \\ \hline
				7.   & gcc-11.2.0-O3         & clang-4.0-O0          & 0.6321     \\ \hline
				8.   & gcc-11.2.0-O3         & clang-5.0-O0          & 0.6323     \\ \hline
				9.   & gcc-6.5.0-Ofast       & clang-4.0-O0          & 0.6378     \\ \hline
				10.  & gcc-6.5.0-Ofast       & clang-6.0-O0          & 0.6379     \\ \hline
			\end{tabular}
			\label{tab:top10_lowest_node_sim}}
	\end{minipage}
	\begin{minipage}[]{0.48\textwidth}
		\centering
		\makeatletter\def\@captype{table}\makeatother\caption{Top 10 lowest neighbor similarities in cross-compiler evaluation}
		\vspace{-5pt}
		\scalebox{0.85}{
			\begin{tabular}{c|c|c|c}
				\hline
				rank & compiler-opt 1             & compiler-opt 2 & neighbor similarity \\ \hline
				1.   & gcc-7.3.0-Ofast              & clang-8.0-O0          & 0.5620     \\ \hline
				2.   & gcc-7.3.0-Ofast               & clang-9.0-O0          & 0.5625     \\ \hline
				3.   & gcc-7.3.0-O3     &     clang-8.0-O0                  & 0.5641     \\ \hline
				4.   & gcc-7.3.0-O3     &         clang-9.0-O0              & 0.5646     \\ \hline
				5.   & gcc-7.3.0-Ofast &       clang-10.0-O0                & 0.5671     \\ \hline
				6.   & gcc-7.3.0-Ofast &    clang-11.0-O0                   & 0.5673     \\ \hline
				7.   & gcc-7.3.0-O3    &    clang-10.0-O0                   & 0.5693     \\ \hline
				8.   & gcc-7.3.0-O3    &      clang-11.0-O0                 & 0.5694     \\ \hline
				9.   & gcc-4.9.4-Ofast  &         clang-8.0-O0              & 0.5706     \\ \hline
				10.  & gcc-4.9.4-Ofast  &     clang-9.0-O0                  & 0.5711     \\ \hline
			\end{tabular}
			\label{tab:top10_lowest_neighbor_sim}}
	\end{minipage}
	\vspace{10pt}
\end{minipage}

Generally, anchor-based methods face the greatest challenge when comparing FCGs compiled by GCC using high optimizations (such as O3 and Ofast) with FCGs compiled by Clang using O0. The anchor similarity can be as low as 0.62, and the neighbor similarity can be 0.56, indicating only about half of the nodes in FCGs or around the anchor functions may have the same name, not even necessarily sharing the identical semantics (33\% are root-equivalent mappings, as shown in Figure~\ref{fig:cross_compiler}). As a result, the anchor-based methods meet a much greater challenge in cross-compiler detection compared to the cross-optimization detection.

\textbf{Observation 2:} The findings from RQ1 further support the observations made in RQ2. The oracle-derived ground truth in RQ1 shows that function mappings are significantly fragmented across different compiler-optimization pairs, which directly impacts the stability of anchors. The instability of anchors, as observed in RQ2, is a direct consequence of the fragmentation of function mappings highlighted in RQ1.

\textbf{Answer to RQ2:} Our oracle exposes that existing methods anchor on fundamentally unstable functions: 62\% of functions in cross-compiler evaluation can serve as boundary functions, of which only 2/3 of presumed anchors are semantically identical (identical mappings). Neighbor similarities can fall below 60\%. This oracle-grounded measurement reveals why anchor propagation fails.

\subsection{RQ3: How does our oracle help reveal the failure modes of clustering-based methods?}
\label{sec:rq3}

In this section, we first use the traditional measurements to evaluate existing clustering-based methods and then use our oracle to reveal the failure modes of existing clustering-based methods.

\subsubsection{Traditional Measurements} 
\begin{figure}[h]
	\centering
	\subfigure[ModX - Clang]{
		\begin{minipage}[t]{0.48\linewidth}
			\centering
			\includegraphics[width=0.95\textwidth]{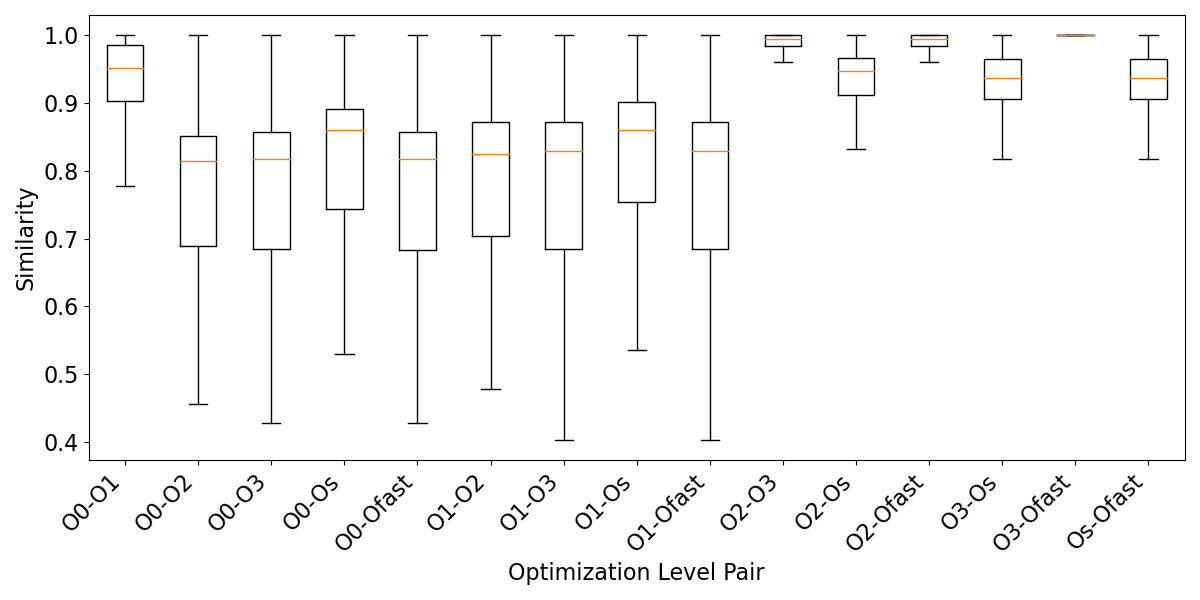}
			\vspace{-15pt}
			\label{fig:modx_cross_opt_clang}
		\end{minipage}%
	}
	\subfigure[ModX - GCC]{
		\begin{minipage}[t]{0.48\linewidth}
			\centering
			\includegraphics[width=0.95\textwidth]{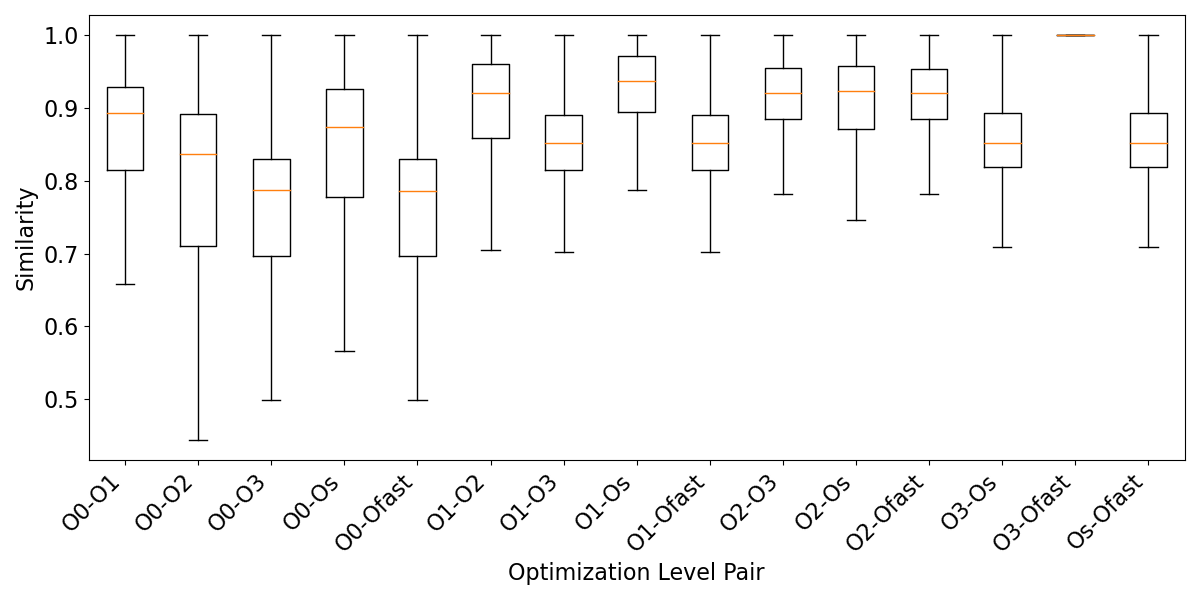}
			\vspace{-15pt}
			\label{fig:gcc_cross_opt_clang}
		\end{minipage}%
	}
	
	\centering
	\subfigure[BMVul - Clang]{
		\begin{minipage}[t]{0.48\linewidth}
			\centering
			\includegraphics[width=0.95\textwidth]{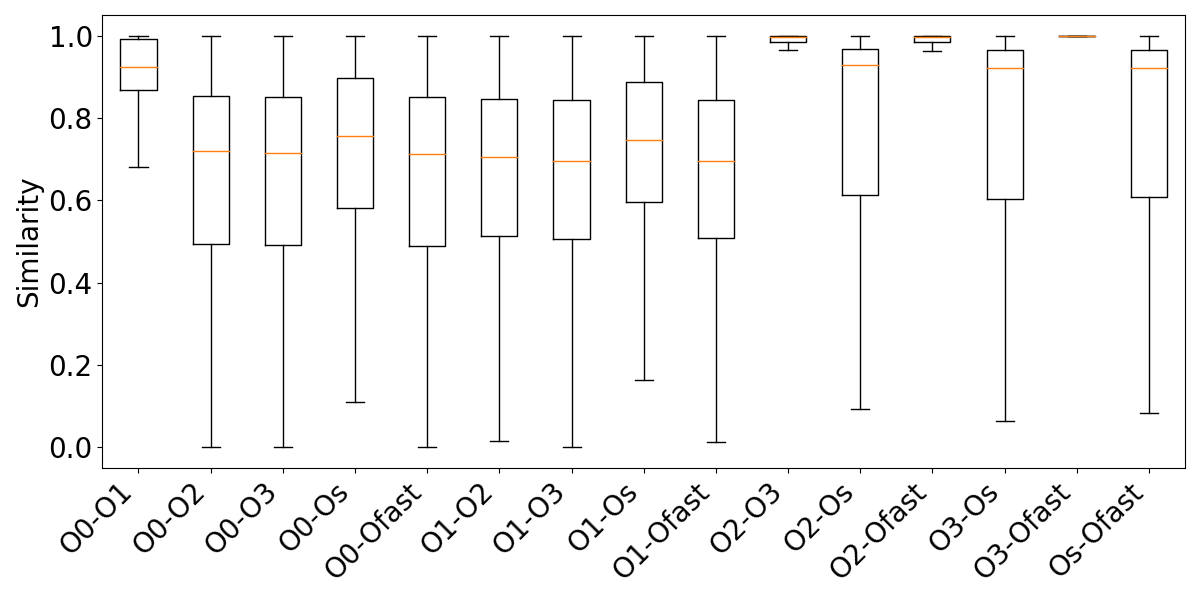}
			\vspace{-15pt}
			\label{fig:bmvul_cross_opt_clang}
		\end{minipage}%
	}%
	\subfigure[BMVul - GCC]{
		\begin{minipage}[t]{0.48\linewidth}
			\centering
			\includegraphics[width=0.95\textwidth]{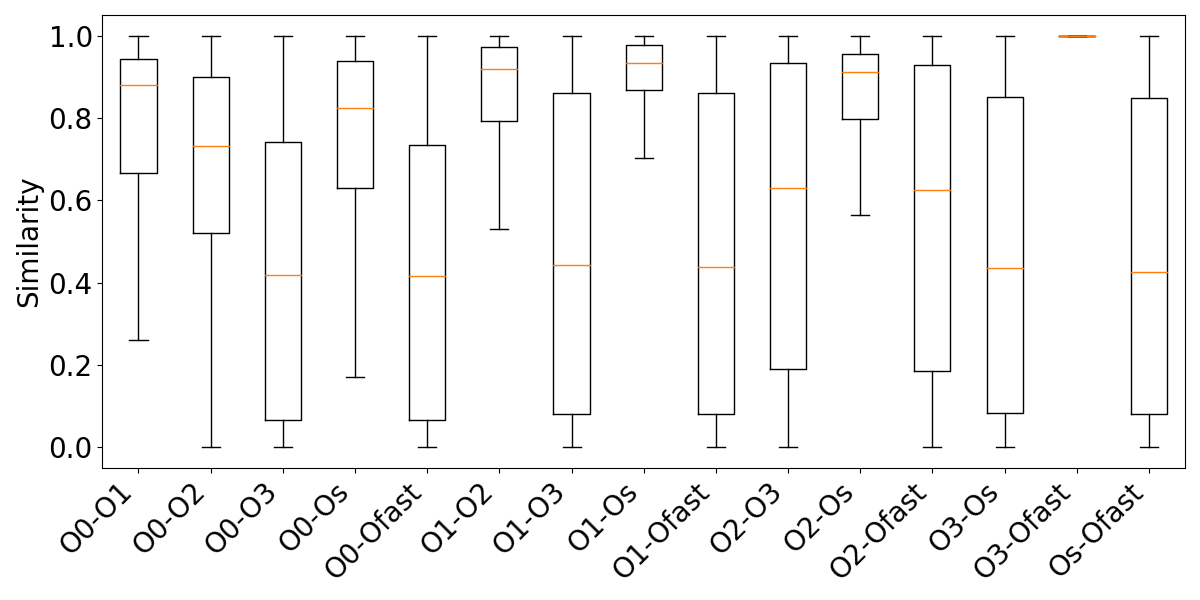}
			\vspace{-15pt}
			\label{fig:bmvul_cross_opt_gccg}
		\end{minipage}%
	}%
	\vspace{-10pt}
	\caption{Cross-optimization evaluation of clustering-based methods}
	\label{fig:clustering_cross_opt}
	\vspace{-10pt}
\end{figure}

\begin{figure}[h]
	\centering
	\subfigure[ModX]{
		\begin{minipage}[t]{0.48\linewidth}
			\centering
			\includegraphics[width=1\textwidth]{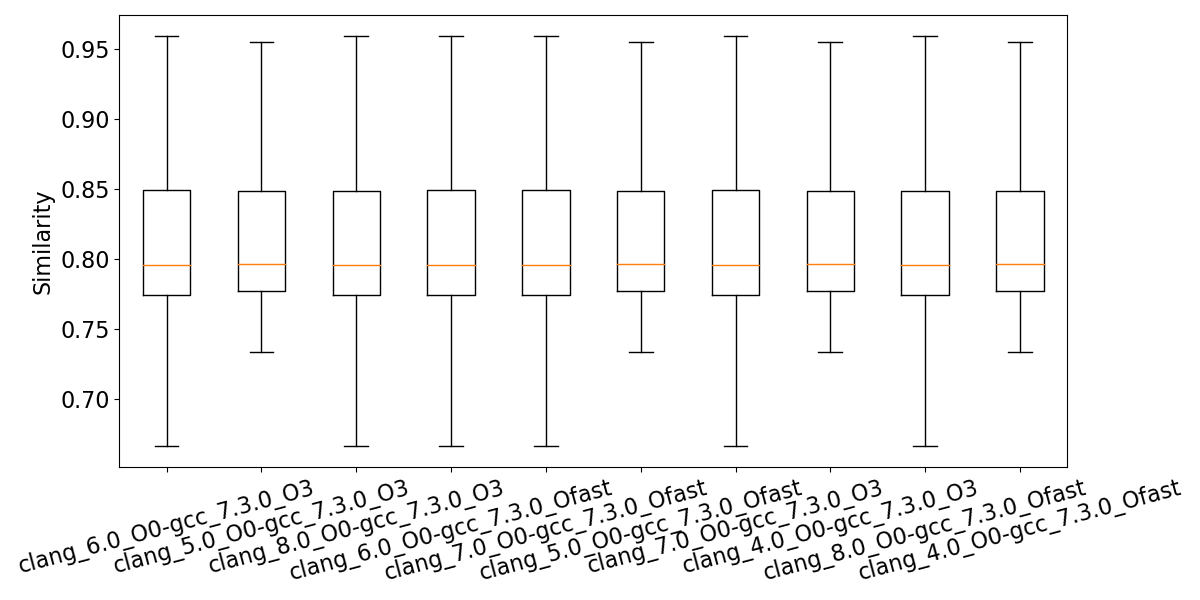}
			\vspace{-15pt}
			\label{fig:modx_cross_compiler}
		\end{minipage}%
	}
	\subfigure[BMVul]{
		\begin{minipage}[t]{0.48\linewidth}
			\centering
			\includegraphics[width=1\textwidth]{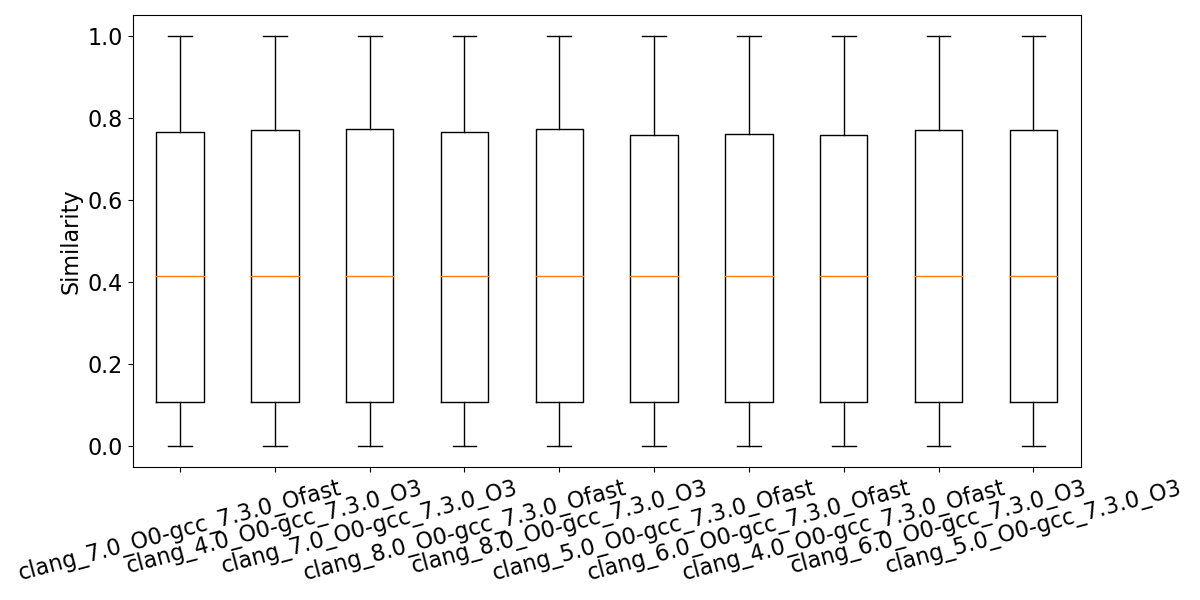}
			\vspace{-15pt}
			\label{fig:bmvul_cross_compiler}
		\end{minipage}%
	}%
	\vspace{-10pt}
	\caption{Cross-compiler evaluation of clustering-based methods}
	\label{fig:clustering_cross_compiler}
\end{figure}

To evaluate ModX and BMVul, we compute the community similarity via Equation~\ref{eq:similarity}  (replace the comparison between MEFR and generated community with the comparison between two generated communities) for the communities generated by ModX and BMVul. Then we use the Equation~\ref{eq:coverage} to find the best matches between the communities, and use the similarities of these best matches to measure how well ModX and BMVul generate semantically equivalent communities. Figure~\ref{fig:clustering_cross_opt} shows the evaluation results of ModX and BMVul in the cross-optimization evaluation. Figure~\ref{fig:clustering_cross_compiler} shows the evaluation results of ModX and BMVul in the 10 most challenging cross-compiler cases (same as Table~\ref{tab:top10_lowest_node_sim} and Table~\ref{tab:top10_lowest_neighbor_sim}. )

Generally, ModX can generate communities with an average similarity that ranges from 70\% to 100\%, presenting a stable distribution of similarities in the evaluation. The average similarity of communities generated by BMVul ranges from 40\% to 100\%, presenting a fluctuating distribution of similarities in the evaluation.

\textbf{Observation 3.1:} The stability of function mappings, as revealed by our oracle in RQ1, has a direct impact on the performance of clustering-based methods. When mappings are stable and maintain a high degree of semantic equivalence across different compilation settings, clustering methods are more likely to generate communities that reflect the true structure of the codebase. However, when mappings are fragmented due to aggressive compiler optimizations, as observed in RQ1, the challenge for clustering methods increases significantly.

\textbf{Observation 3.2:} Traditional measurements, which focus solely on surface-level similarity distributions, are insufficient to capture the nuances of the generated communities. These methods fail to consider the minimal region of function mappings with equivalent semantics—a critical aspect for accurately assessing community quality. The inability to account for the minimal region means that these metrics might overestimate the performance of clustering methods, as the clustering methods may over-aggregate boundary functions, which will also increase the community similarities.

\subsubsection{Oracle Measurements}

\begin{figure*}[h]
	\centering
	\subfigure[ModX]{
		\begin{minipage}[t]{0.32\linewidth}
			\centering
			\includegraphics[width=1\textwidth]{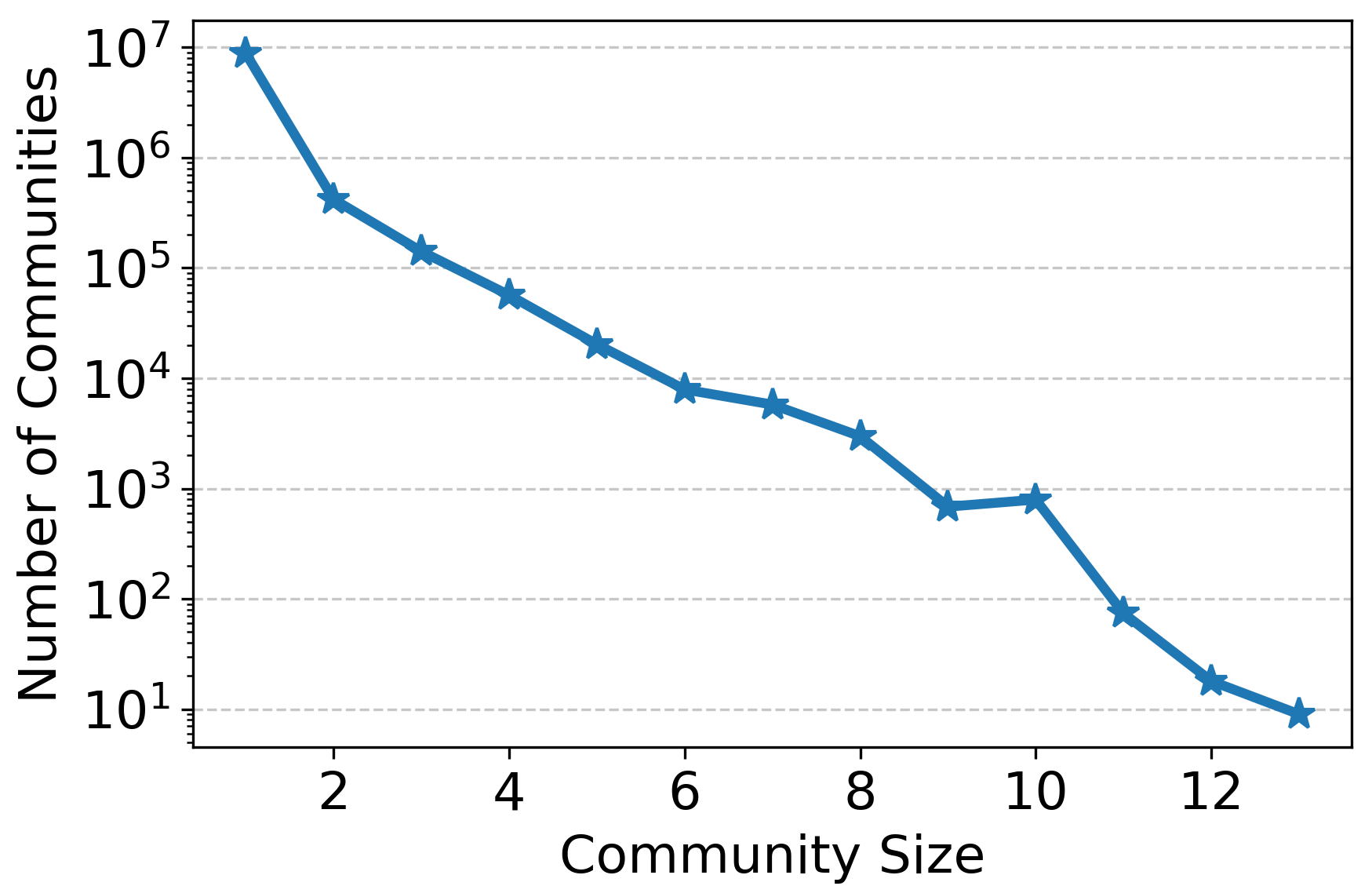}
			\vspace{-15pt}
			\label{fig:modx_size}
		\end{minipage}%
	}
	\subfigure[BMVul]{
		\begin{minipage}[t]{0.32\linewidth}
			\centering
			\includegraphics[width=1\textwidth]{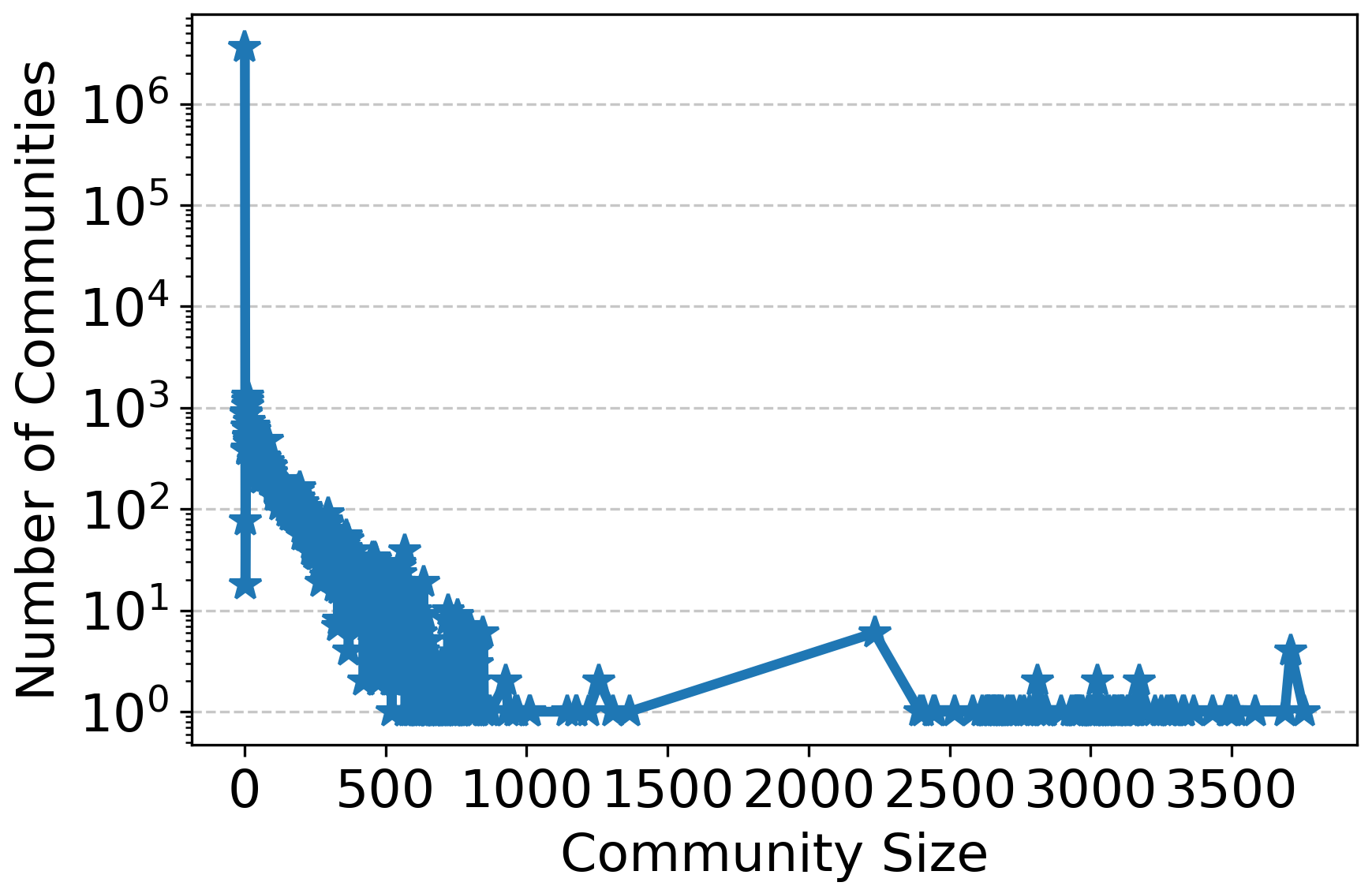}
			\vspace{-15pt}
			\label{fig:BMVul_size}
		\end{minipage}%
	}%
	\subfigure[Oracle]{
		\begin{minipage}[t]{0.32\linewidth}
			\centering
			\includegraphics[width=1\textwidth]{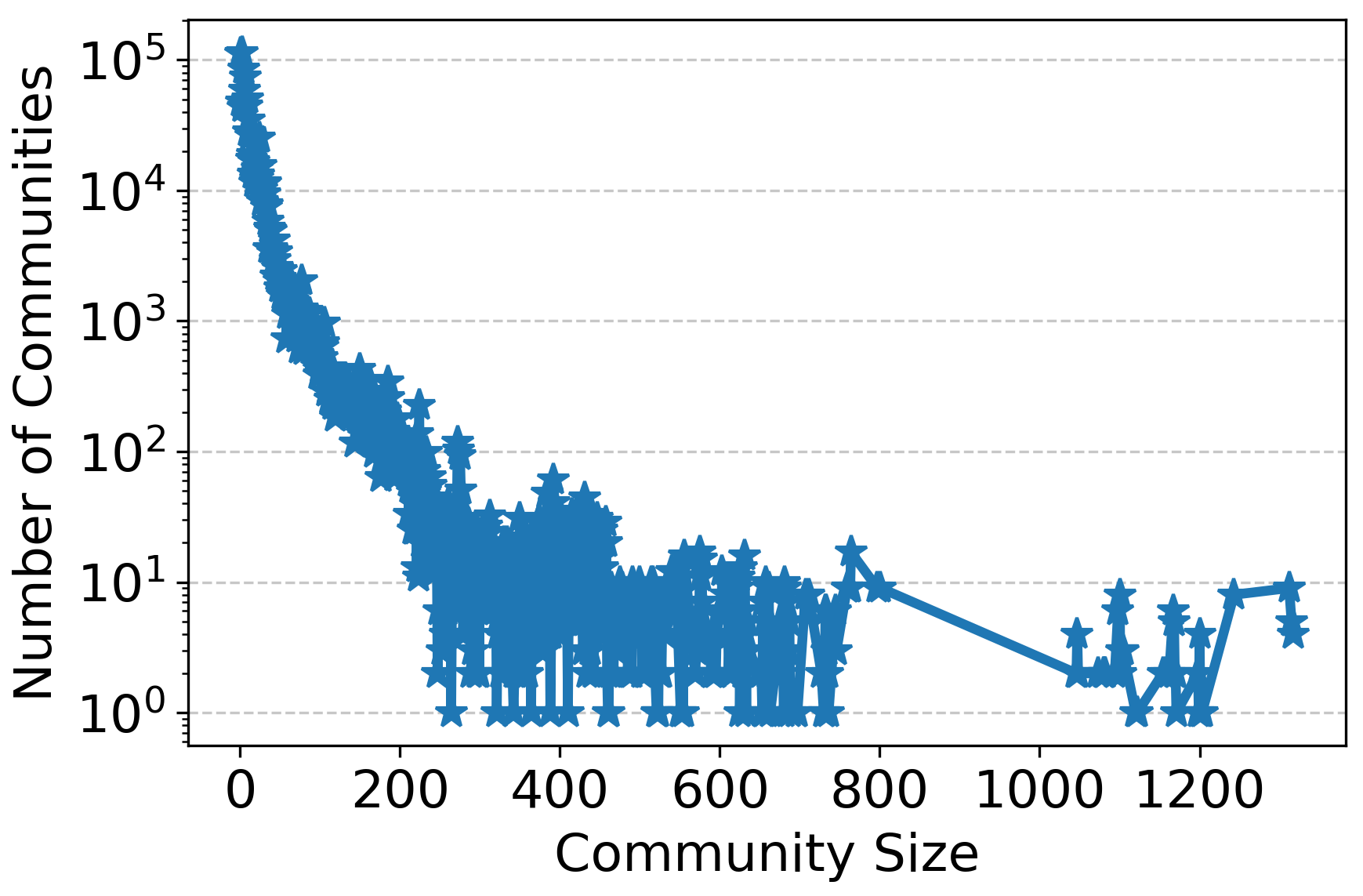}
			\vspace{-15pt}
			\label{fig:anchor_size}
		\end{minipage}%
	}%
	\vspace{-10pt}
	\caption{Community size distributions in ModX, BMVul, and Oracle}
	\label{fig:com_size}
	\vspace{-10pt}
\end{figure*}

\begin{figure}[h]
	\centering
	\subfigure[ModX vs. Oracle]{
		\begin{minipage}[t]{0.48\linewidth}
			\centering
			\includegraphics[width=0.85\textwidth]{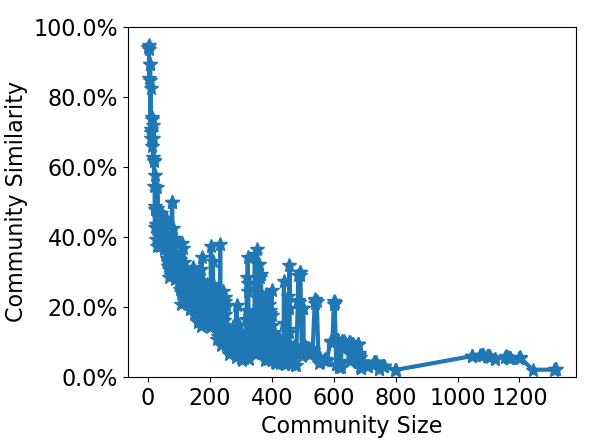}
			\vspace{-15pt}
			\label{fig:modx_sim_to_optimal}
		\end{minipage}%
	}	
	\subfigure[BMVul vs. Oracle]{
		\begin{minipage}[t]{0.48\linewidth}
			\centering
			\includegraphics[width=0.85\textwidth]{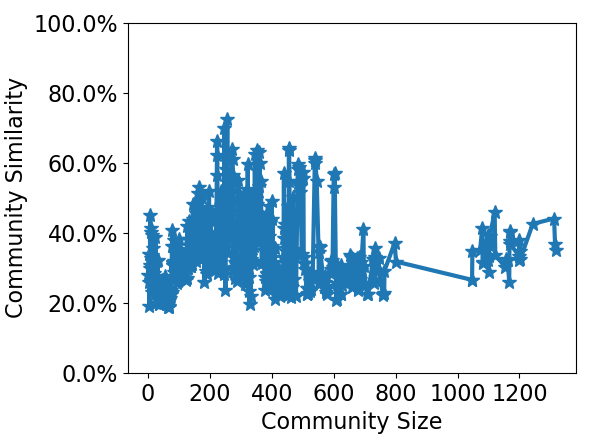}
			\vspace{-15pt}
			\label{fig:BMVul_sim_to_optimal}
		\end{minipage}%
	}%
	\vspace{-10pt}
	\caption{Oracle-measured community similarities}
	\label{fig:com_sim_to_optimizal}
\end{figure}

\begin{figure}[h]
	\centering
	\subfigure[ModX under-aggregation]{
		\begin{minipage}[t]{0.48\linewidth}
			\centering
			\includegraphics[width=0.85\textwidth]{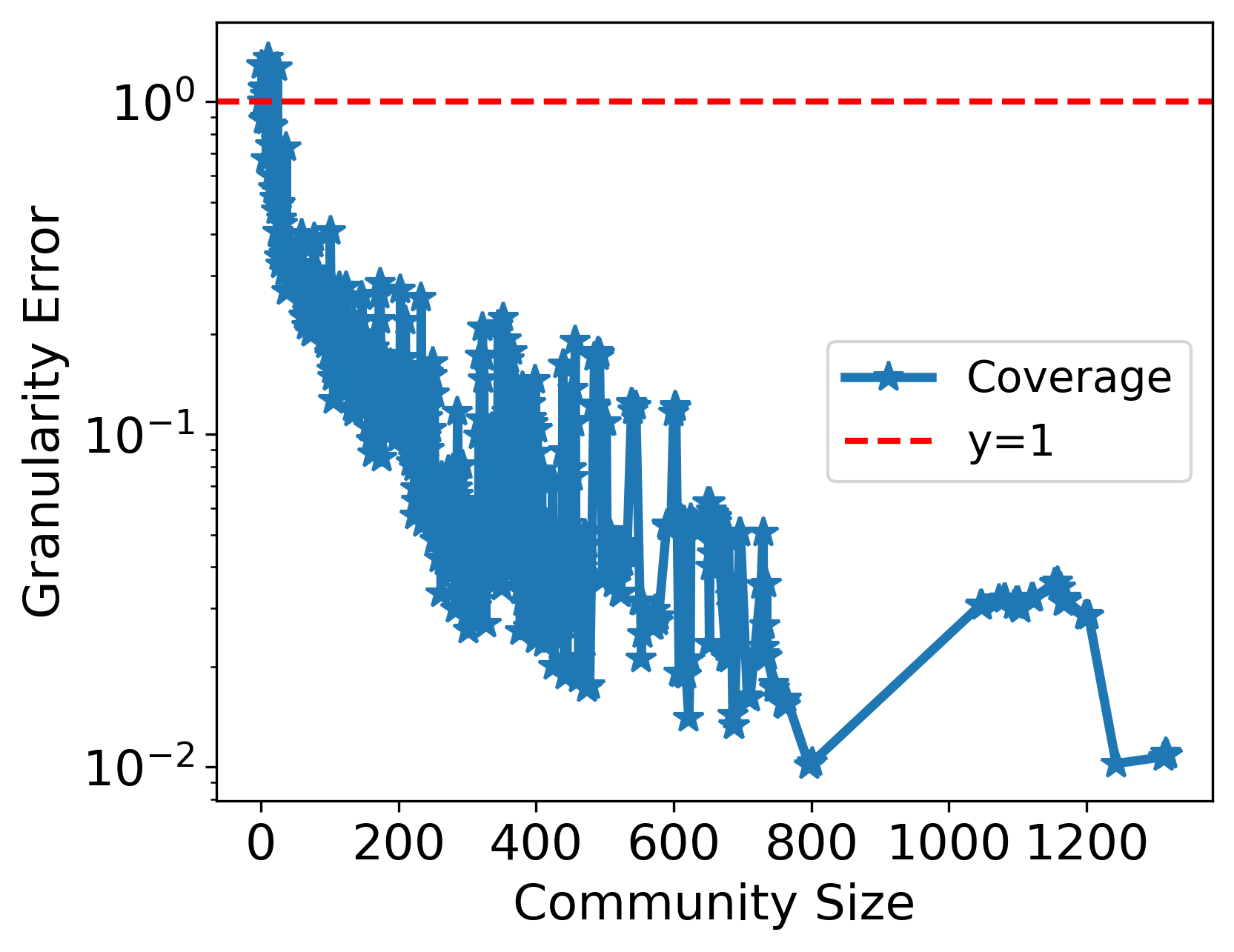}
			\vspace{-15pt}
			\label{fig:modx_coverage}
		\end{minipage}%
	}	
	\subfigure[BMVul over-aggregation]{
		\begin{minipage}[t]{0.48\linewidth}
			\centering
			\includegraphics[width=0.85\textwidth]{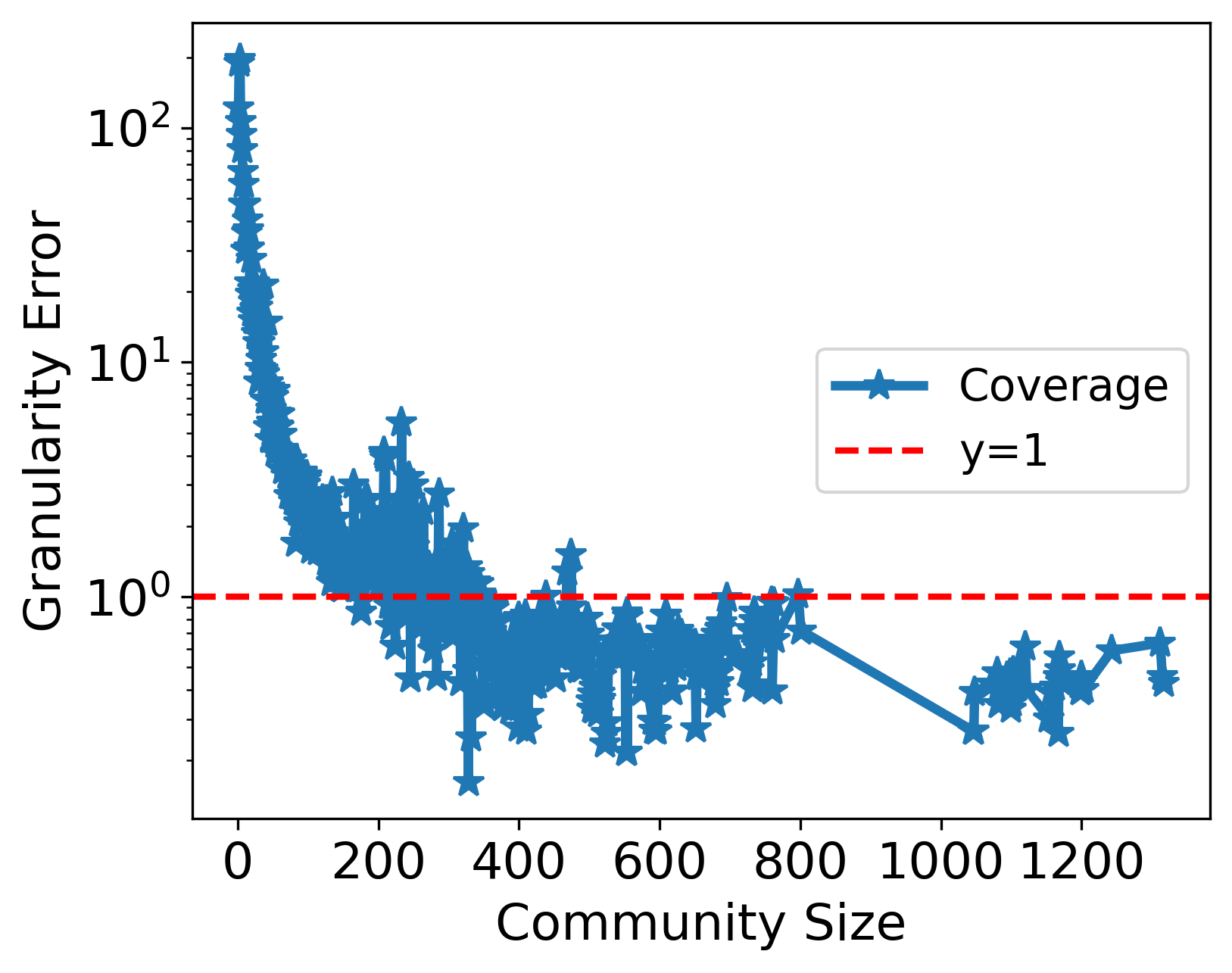}
			\vspace{-15pt}
			\label{fig:BMVul_coverage}
		\end{minipage}%
	}%
	\vspace{-10pt}
	\caption{Oracle-measured granularity error}
	\label{fig:coverage}
\end{figure}

Figure~\ref{fig:com_size} shows the distribution of community size in ModX, BMVul, and our oracle (MEFR). Figure~\ref{fig:com_sim_to_optimizal} shows the distribution of community similarity of ModX and BMVul when compared to our oracle using Equation~\ref{eq:coverage}. Figure~\ref{fig:coverage} shows the distribution of granularity error of ModX and BMVul when compared to our oracle using Equation~\ref{eq:gran}. The analysis results are as follows.

\textbf{ModX:}
As shown in Figure~\ref{fig:modx_size}, though ModX can achieve an average similarity of 80\% when conducting the cross-compiler evaluation, the community size of ModX only ranges from 1 to 13, while the community size of the MEFR will range from 1 to 1317, indicating a low coverage of ModX when generating communities for binary functions with aggressive inlining.

As a result, as shown in Figure~\ref{fig:modx_sim_to_optimal}, the community similarity decreases dramatically when the community size increases. When the community size is larger than 200, the communities generated by ModX and the MEFR only share less than 40\% of source functions.

Figure~\ref{fig:modx_coverage} further verifies the under-aggregation of ModX. As shown in Figure~\ref{fig:modx_coverage}, as the community size grows, the communities generated by ModX can only cover about 10\% of the source functions of MEFR. These statistics indicate that ModX tends to generate smaller communities that can not fully capture the semantically equivalent regions.

As observed in our dataset, about 3\% of binary functions compiled by O3 have inlined more than 13 source functions. When comparing these BFIs with NBFs generated by O0, at least 14 NBFs should compose a community to match these BFIs. However, the biggest community generated by ModX only has 13 members, which means that it cannot cover these binary functions. Moreover, as studied in \cite{jia20231}, a binary function in OpenSSL~\cite{openssl} can inline 2000 source functions. ModX is completely lacking in its ability to detect these functions.

In summary, although the communities generated by ModX are closer to the optimal decomposition when the community size is less than 13, it cannot adapt to the binary functions that have inlined a large number of source functions.
The size of communities may limit the ability of ModX to match binary functions with a large number of inlined source functions, making ModX less scalable to the binaries enabled with aggressive inlining.

\textbf{BMVul:}
As shown in Figure~\ref{fig:BMVul_size}, generally, the size of communities decomposed by BMVul can range from 1 to 3489, indicating that BMVul applies a more aggressive strategy. However, the community size of MEFR only ranges from 1 to 1317, indicating that the clustering strategy of BMVul is too aggressive, which may include unnecessary functions in the communities.

As a cost, the aggressive strategies that BMVul applied resulted in unstable community similarities, as shown in Figure~\ref{fig:BMVul_sim_to_optimal}.  The community similarity mainly ranges from 20\% to 60\%. Even for the community composed of only one function, which may not be influenced by inlining, the communities generated by BMVul only share less than 40\% of source functions with the MEFR.

Figure~\ref{fig:BMVul_coverage} also presents the aggressive strategies of BMVul. For a community containing one function in the oracle, BMVul tends to generate a community containing more than one hundred functions. The aggressive strategies of BMVul result in its low community similarities.

\subsubsection{Failure Modes}

Using the optimal decomposition identified in our work, we can provide a more comprehensive view of why existing works cannot resolve the binary decomposition under FCG variance. We hope our findings could inspire subsequent works to propose more reliable and effective methods.

Our oracle enables precise diagnosis by comparing community size distributions and measuring deviation from optimal decomposition, revealing two distinct failure modes.

\textbf{Observation 3.3: Oracle exposes ModX's under-aggregation failure.} ModX's maximum community size (13) is far below the size of MEFR in oracle (1,317). The oracle measured similarities drop below 40\% for communities containing >200 functions, revealing ModX cannot handle aggressive inlining.

\textbf{Observation 3.4: Oracle exposes BMVul's over-aggregation failure.} The oracle shows BMVul generates communities up to 3,489 functions, far exceeding oracle optimal size (1,317). This includes semantically unrelated functions, causing unstable oracle similarity scores.

\textbf{Answer to RQ3:} Our oracle precisely diagnoses complementary failure modes: ModX's conservative granularity causes under-aggregation (missing MEFRs), while BMVul's aggressive expansion causes over-aggregation (merging MEFRs). The oracle reveals neither method achieves optimal decomposition.

\section{Discussion}

\subsection{Oracle Construction}

\textbf{Practicality and Limitations.} Our oracle construction methodology represents a pragmatic solution employing static analysis and debug symbols to establish ground-truth for binary decomposition. This approach provides reproducible results and scales to large datasets, making it suitable for systematic evaluation. However, it is essential to recognize its inherent limitations: the reliance on debug information (\texttt{-g} flag) restricts its applicability to production binaries where such metadata is typically stripped. Moreover, our current focus on function inlining as the primary source of FCG variance, while justified by empirical evidence, may not capture all compilation-induced transformations such as function outlining.

\textbf{Alternative Oracle Construction Paradigms.} In real-world scenarios where debug symbols are unavailable, alternative oracle construction methods become necessary. One practical approach involves manual annotation by domain experts for a representative subset of binaries, establishing high-confidence ground-truth that can serve as validation benchmarks for automated methods. Another direction is dynamic execution-based oracles, which trace function call sequences at runtime across different compilation settings to capture behavioral variations that static analysis might miss. While such methods face scalability challenges, they could complement our static approach for specific critical functions.

\textbf{Future Extension.} Our research can be extended in several directions. First, adapting the methodology to account for additional compiler transformations—such as function outlining—would create a more comprehensive evaluation framework. Second, integrating dynamic analysis techniques could create hybrid oracles that combine the scalability of static analysis with the precision of runtime observations, particularly for detecting unobvious compiler optimizations and inter-procedural control flow variations. Third, exploring the applicability of our methodology to other binary analysis domains—such as vulnerability detection, malware classification, or reverse engineering assistance—could broaden its impact beyond binary decomposition. By establishing robust evaluation frameworks in these adjacent fields, we can enable more rigorous validation of tools that rely on accurate binary understanding. Furthermore, investigating the interaction between different compiler optimization levels and binary obfuscation techniques would provide deeper insights into the robustness of analysis methods under adversarial conditions.

\subsection{Evaluation Metrics}

\textbf{Significance of Oracle-Grounded Absolute Metrics.} We propose the first oracle-based metric suite that quantifies \emph{absolute} rather than relative decomposition quality. All scores are computed against the MEFR derived in \S\ref{sec:Oracle_Construction}, enabling us to pinpoint whether failures stem from algorithmic defects or fundamental information loss induced by inlining. The three core metrics --- $S_{\text{anch}}$, $S_{\text{nb}}$, and $\mathcal{G}(M)$ --- each target specific failure modes: anchor loss, structure loss, and granularity mismatch. These metrics provide actionable insights for algorithm improvement.

\textbf{Limitations of Current Metrics.} Despite their effectiveness, our metrics have inherent limitations. The metrics for anchor-based methods, including  $S_{\text{anch}}$ and $S_{\text{nb}}$, currently treat all functions equally, without considering the varying size of different functions. This uniform weighting may also overlook the semantic difference of functions, especially functions implemented for complex functionalities. Similarly, $\mathcal{G}(M)$ also treats source functions equally, which may not accurately assess the semantic similarity of binary functions.

\textbf{Generalizability to Other Domains.} Our metrics' principles---anchor stability, neighbor stability, and granularity-aware scoring---can be extended to CFG matching, malware lineage analysis, and vulnerability detection. We hope our evaluation metrics can inspire researchers to find more insights hidden in the binary code analysis.

\subsection{Threats to Validity}

\textbf{Internal Validity.} The use of IDA Pro for FCG construction introduces potential disassembly errors, which may propagate to the labeling process and affect results. While IDA Pro is a state-of-the-art commercial tool widely used in academia and industry \cite{deepbindiff, pang2021sok}, no disassembler guarantees perfect accuracy, particularly for heavily optimized or obfuscated binaries.

\textbf{External Validity.} Our evaluation covers 17 compilers and 6 optimizations across 4 architectures, but cannot capture all possible compilation settings (e.g., non-default optimizations \cite{10.1145/3453483.3454035}). The inlining patterns observed may not generalize to projects with highly customized inlining conventions. Nevertheless, our labeling method is designed to be portable and can be applied to other datasets as long as inlining related information can be extracted.

\textbf{Construct Validity.} Our reliance on third-party tools (IDA Pro, DWARF debug tools) introduces potential inaccuracies in function inlining identification \cite{jia20231}. We mitigate this by selecting the most reliable tools available and validating our oracle through manual inspection of 100 random FCG pairs, confirming semantic equivalence and minimality of the generated MEFRs.

\section{Conclusion}

In this work, we establish the oracle for binary decomposition under compilation variance. We construct precise FCG mappings across 17 compilers, 6 optimizations, and 4 architectures, and identify minimum semantic-equivalent function regions as ground truth. This enables the first rigorous evaluation of decomposition algorithms, revealing critical limitations: existing methods suffer from either under-aggregation or over-aggregation.

In the future, we plan to extend this work along three primary directions. First, we will develop \textit{compilation-aware decomposition algorithms} that leverage the MEFR structure to dynamically adjust aggregation strategies, thereby mitigating the under-/over-aggregation failures identified in current methods. Second, we aim to generalize the oracle construction to handle more aggressive compiler transformations beyond inlining, such as function splitting, tail-call elimination, and interprocedural optimization, which challenge the stability of boundary functions. Finally, we intend to release our oracle construction toolkit as a public benchmark, enabling the community to diagnose algorithmic deficiencies and drive the next generation of robust binary analysis tools.

\bibliographystyle{ACM-Reference-Format}
\bibliography{reference}
	
\end{document}